\documentclass[acmsmall]{acmart}

\AtBeginDocument{%
  \providecommand\BibTeX{{%
    \normalfont B\kern-0.5em{\scshape i\kern-0.25em b}\kern-0.8em\TeX}}}

\setcopyright{acmcopyright}
\copyrightyear{2025}
\acmYear{2027}
\acmDOI{XXXXXXX.XXXXXXX}

\acmJournal{JACM}
\acmVolume{37}
\acmNumber{4}
\acmArticle{111}
\acmMonth{8}

\acmPrice{15.00}
\acmISBN{978-1-4503-XXXX-X/18/06}

\usepackage{subcaption}
\usepackage{threeparttable}
\usepackage{tabularx}
\usepackage{longtable}
\usepackage{graphicx}
\usepackage{array}
\usepackage{booktabs}
\usepackage{tcolorbox}
\usepackage{multirow}

\usepackage{amssymb}
\usepackage{tikz}
\usetikzlibrary{arrows.meta,positioning}
\usepackage{adjustbox}
\usepackage{pgfplots}
\pgfplotsset{compat=1.18}
\usepackage{xcolor}
\usepackage{enumitem}

\definecolor{mplblue}{HTML}{1F77B4}
\definecolor{mplorange}{HTML}{FF7F0E}
\definecolor{mplgreen}{HTML}{2CA02C}
\definecolor{mplred}{HTML}{D62728}
\definecolor{mplpurple}{HTML}{9467BD}
\definecolor{mplbrown}{HTML}{8C564B}

\usepackage{tikz}
\usetikzlibrary{arrows.meta,positioning}

\usepackage{xcolor}

\definecolor{insta}{HTML}{8E44AD}
\definecolor{youtube}{HTML}{D62728}
\definecolor{tiktok}{HTML}{2CA02C}
\definecolor{axisgray}{HTML}{A8A8A8}
\definecolor{textgray}{HTML}{333333}

\begin{document}

\newcommand{\subsubsubsection}[1]{\textbf{#1.}}

\title{
Reply, Delete, or Ignore? Examining How Content Creators Perceive and Select Comment Moderation Strategies
}

\author{Yunhee Shim}
\email{yunhee.shim@rutgers.edu}
\affiliation{%
  \institution{Rutgers University}
  \city{New Brunswick, NJ}
  \country{USA}
  }

\author{Shagun Jhaver}
\email{shagun.jhaver@rutgers.edu}
\affiliation{%
  \institution{Rutgers University}
  \city{New Brunswick, NJ}
  \country{USA}
  }

\renewcommand{\shortauthors}{Shim and Jhaver}

\begin{abstract}

Content creators on social media sites occupy highly visible positions on their channels. As a result, creators, especially those with large followings, experience disproportionate levels of online harm. 
To address such harm, they enact a range of moderation strategies, which in turn shape the visibility of content that their audiences encounter.
This paper examines how content creators perceive three moderation strategies to address hateful comments---deleting, replying to, or simply ignoring---and how they decide which strategy to deploy.
While creator moderation is usually examined through the lens of safety, creators’ regulation decisions may also be shaped by concerns about how their actions appear to audiences and how they are rewarded or penalized by platforms' recommendation algorithms. 
Conducting a survey of 584 content creators, we found
that in their view, (1) deleting is the most beneficial for achieving safety, (2) both deleting and replying produce better impression management benefits than ignoring, and (3) replying is perceived to yield the highest algorithmic benefits. 
Crucially, while expectations of emotional safety and impression management benefits significantly predicted creators' willingness to adopt each comment moderation strategy, perceived algorithmic benefits did not.
By unpacking how creators evaluate these trade-offs, this study contributes to HCI research on understanding creator-led, middle-level governance.
We conclude with design implications for supporting creators as crucial governance actors without burdening them with sole responsibility for online safety.
\end{abstract}



\maketitle

\section{Introduction}
From content publishing to real-time livestreaming, content creators frequently receive high public visibilty on their channels, often at the cost of exposure to online harassment~\cite{eckert2018fighting,samermit2023millions}, ranging from criticism to severe threats.
Compared to non-creators, creators face a disproportionately greater risk of encountering such harms~\cite{thomas2022s,matthews2025supporting}.
To address these harms, they adopt a range of moderation strategies, including platform-offered content removal tools (e.g., reporting~\cite{crawford2016flag}, and deleting~\cite{thomas2022s}), visibility limiting tools (e.g., word filters~\cite{jhaver2022designing} and toxicity sliders~\cite{jhaver2023personalizing}), and discursive strategies (e.g., counterspeech~\cite{atreja2023remove,huang2025empowering} and public humiliation~\cite{shim2025pin,corry2021screenshot}).

Through these practices, creators protect themselves from immediate harm while also shaping how, and whether, harmful comments remain visible to their audiences.
These creator-led moderation responses to online harm can enforce the norms within creators' own channels, define the audience interpretation of harmful interactions, and encourage desired forms of participation~\cite{jhaver2023decentralizing}. 
More broadly, creator-led harm addressing practices can validate the victims' experiences~\cite{blackwell2017classification}, and serve as prominent markers of pushbacks against threats that can empower audiences~\cite{geiger2016bot,jhaver2022designing,shim2025pin}.
In this sense, creators can be understood as middle-level governance actors whose moderation practices extend beyond ensuring merely their own individual safety~\cite{jhaver2023decentralizing}. 

Given creators' crucial role in platform governance, HCI scholars have examined the types of harms that creators encounter and the strategies they use to address those harms~\cite{soneji2024feel,samermit2023millions,stegeman2024strategic,shim2025pin,heung2024vulnerable}.
This body of work has offered valuable insights for designing moderation tools that better support creators. 
However, we do not yet fully understand how creators perceive, compare, and evaluate different moderation strategies when deciding whether to adopt them.
Because platform-provided moderation tools are often designed to maintain safe online communities and prevent further harms~\cite{cai2024content,goldman2021content}, it is important to examine safety as one of the expected outcomes of moderation practices and assess whether these expectations are met and how they shape moderation choices.
Yet, this focus raises a critical question: what other consideration besides safety shape the ways that creators evaluate and adopt different comment moderation options? 

We examine creators’ moderation strategies by situating them as key participants within platforms' ``attention economies~\cite{goldhaber1997attention},'' where engagement functions as a form of currency tied to visibility, audience retention, and sometimes even creators’ livelihood~\cite{bertaglia2021clout}.
Thus, in their everyday social media use, creators may seek to maintain an approachable public image and enhance audience engagement~\cite{marwick2015instafame}, while also appraising how algorithmic systems shape their visibility~\cite{cotter2019playing,bishop2019managing}.
These considerations suggest that creators’ moderation actions, particularly how they manage audience comments on their content, may be shaped not only by safety concerns but also by their goals tied to maintaining a desired public impression and algorithmic visibility.


Thus, besides safety, we conceptualize impression management as a second dimension that impacts creators' perceptions of harm-addressing mechanisms and their moderation decisions.
Prior research shows that creators’ desire to manage their public image shapes a wide range of their behaviors, from uploading aspirational posts to interacting with audiences in warm and responsive ways~\cite{rahmatulah2024impression,sukmayadi2024constructing}.
Examining impression management as a dimension therefore helps reveal how audience facing concerns influence creators' moderation perceptions and choices.

Further, we examine incentives tied to recommendation algorithms as the third dimension through which creators evaluate moderation strategies.
Because platforms' recommendation algorithms substantially shape creators' visibility and monetization~\cite{hodl2023content,lee2025managing,devito2022transfeminine}, creators often try to infer how these systems work and adjust their practices such as choosing content topics or scheduling posting activities in ways they believe will receive greater algorithmic visibility~\cite{wu2019agent,echauri2026algorithmic}.
Given the growing evidence on how algorithmic considerations influence a wide range of work practices and behavioral patterns~\cite{jhaver2018algorithmic,glotfelter2019algorithmic,guess2023social,cotter2019playing}, we examine how such considerations affect creators' moderation work.

Bringing these three dimensions together, we examine how various harm-addressing strategies speak to creators' concerns about safety, impression management, and algorithmic visibility, and how these considerations collectively shape creators’ moderation decisions.
Accordingly, we pose the following research questions:

\vspace{5pt}
\textit{\textbf{RQ1}: How do content creators compare different comment moderation strategies across the dimensions of perceived safety, impression management, and algorithmic visibility?}

\textit{\textbf{RQ2}: How do creators' expectations regarding the influence of each moderation strategy on safety, impression management, and recommendation algorithms influence their adoption of that strategy?}
\vspace{5pt}

We explore three moderation strategies---replying to, deleting, and ignoring---in this study that vary in their consequences, moderation intensity, and required effort, allowing us to compare how creators evaluate moderation actions across different dimensions. 
We elaborate on this rationale in Section ~\ref{relatedwork_2.1}.
To examine these questions in a concrete moderation context, we focus on creators’ responses to hate speech comments. 
Hate speech is one of the most prevalent forms of online harms~\cite{pater2016characterizations,wilson2020hate,weerasinghe2025beyond}. Identity-based attacks, in particular, can extend beyond direct targets, by implicating the broader communities invoked by such  comments~\cite{chatzakou2017measuring,pater2016characterizations,thomas2021sok,scheuerman2021framework,samermit2023millions}.
Because hate speech often operates through identity-based denigration and is closely associated with emotional harm~\cite{samermit2023millions,scheuerman2021framework}, it provides a concrete context in which creators can evaluate and self-report the emotional safety associated with different moderation responses, alongside the other considerations examined in this study.
We therefore operationalized online harm through an identity-based hate speech scenario, allowing participants to evaluate moderation strategies in a realistic, everyday context.
Using this scenario, we conducted an online survey with content creators, recruited through Prolific (N = 584).
The survey used a within-subjects design, where participants evaluated each moderation strategy in terms of its perceived benefits for safety, impression management, and recommendation algorithms, and their likelihood of adopting each strategy. 

Our survey findings show that creators evaluated deleting, replying to, and ignoring harmful comments differently across the three dimensions. 
Deleting was perceived as the most effective strategy for attaining safety, followed by ignoring and replying. 
For impression management, deleting and replying were perceived as more beneficial than ignoring, while the difference between deleting and replying was not significant. 
Replying was perceived as offering the greatest algorithmic incentives, followed by ignoring and deleting. 
Finally, perceived safety and impression management benefits significantly predicted creators’ likelihood of adopting each of the three moderation strategies, whereas perceived algorithmic incentives did not.

Our study contributes to prior research on creator practices that reduce harm~\cite{soneji2024feel,thomas2022s,heung2024vulnerable,shim2025pin,grober2024chose} and the attention economies in which they operate~\cite{ma2023multi,soneji2024feel,thomas2022s,devito2022transfeminine,duffy2023platform,kopf2020rewarding,stegeman2024strategic}. 
We show that creators’ moderation decisions are audience-facing actions that extend well beyond self-protection. 
Through these actions, creators actively seek to be seen as responsible moderation actors, suggesting that their moderation practices should be understood in relation to their desired roles in middle-level platform governance.
Further, creators evaluate each moderation strategy distinctly across safety, impression, and algorithmic dimensions, highlighting the need to account for their situated understandings of platform visibility dynamics~\cite{cotter2019playing,marwick2015instafame}.

Building on our findings, we advocate for recognizing creators as stewards of their comment spaces and better supporting their middle-level governance activities, thereby contributing to robust multi-level platform moderation~\cite{jhaver2022designing,jhaver2023decentralizing}.
Our strategy-specific insights can inform the design of creator-led moderation tools by illustrating how visibility-reduction (deleting), audience-facing response (replying), and non-engagement (ignoring) uniquely shape creator safety, viewer relationships, and algorithmic labor.

\section{Related Work}
\label{related work}

\subsection{Content Creators’ Adoption of Moderation Strategies}
\label{relatedwork_2.1}
To address inappropriate comments, content creators may adopt various safety measures offered by the platform, such as reporting, deleting, or muting~\cite{heung2024vulnerable,thomas2022s}.
These visibility-control mechanisms mitigate harms by constraining or removing exposure to inappropriate comments~\cite{grimmelmann2015virtues}. 
Thus, these tools enact localized defensive actions, facilitating scalable moderation by distributing governance work across individual creators~\cite{grimmelmann2015virtues,atreja2023remove}.  

Among these approaches, we pay particular attention to \textbf{deletion}, as it enables creators to directly regulate their comment space through platform-sanctioned authority, without requiring platforms' explicit approval for each action.
This creator-enacted deletion of harmful comments is a distinctive localized governance action, as it grants creators more direct and immediate control over their comment space than moderation approaches such as flagging or hiding.
Given that deletion enables site-wide removal, it has the potential to contribute to collective community safety by keeping harmful content away from public view~\cite{crawford2016flag,thomas2022s}.
However, deletion may provoke backlash or public criticism of censorship~\cite {john2024classification,myers2018censored}.

In addition to adopting platform-offered moderation tools, creators may also intervene in online harms in a discursive way---this occurs most directly through \textbf{replying to} inappropriate comments.
These replies can take various rhetorical forms, including warning, shaming, or responding with sarcasm or empathy~\cite{ziegele2020not,shim2025pin}.
Such responses function as counterspeech, i.e., they publicly contest inappropriate behavior~\cite{atreja2023remove,buerger2022they}, and signal normative boundaries within localized comment spaces~\cite{rieger2018hate,kalch2017replying}.
Counterspeech empowers the community members to confront inappropriate comments and, when sustained over time, can reinforce prosocial norms~\cite{shim2025pin,murumaa2021misogynist,im2022women}.
For instance, it may prompt commenters and observers to change their behavior; in some cases, commenters delete or apologize for their inappropriate comments~\cite{wright2017vectors,shim2025pin,schieb2016governing}.
While replying has these benefits
, it can also heighten the risk of escalation~\cite{rashidi2020s,shim2025pin} and retaliation~\cite{ping2024behind}, which could reinforce a hostile environment~\cite{baider2023accountability}.
Additionally, engaging in counterspeech demands considerable time and effort~\cite{haenlein2020navigating}, compared to simply deleting harmful comments.

We also shed light on \textbf{ignoring} inappropriate comments as a strategic response to online harms. 
We define \textit{ignoring} in this study as creators' deliberate decision not to respond to or engage with harmful commenters in order to avoid amplifying toxicity or escalating confrontation.
Although this strategy is often perceived as passive \textit{non}-reaction, a substantial proportion of creators strategically choose to disengage from online interactions in response to hate, both for immediate emotional relief and for longer-term safety effects~\cite{thomas2022s,samermit2023millions}.
Ignoring can also serve as a community-oriented safety practice by minimizing harm through the refusal to grant disruptive behavior additional visibility, a norm often summarized as \textit{do not feed the trolls}~\cite{kiesler2012regulating}.
In this sense, ignoring may function as a form of moderation through a strategic non-intervention.
While ignoring minimizes effort and avoids direct confrontation, it may allow harmful content to remain visible, potentially exposing broader audiences to harm and failing to signal that such behavior violates community norms.

We focus on these three strategies---deleting, replying, and ignoring---because their effects are immediate, they operate at the level of a specific comment rather than at the account level, and they are fully under creators' control.
Other moderation practices, such as blocking, muting, reporting, filtering, or hiding, may depend more heavily on platform agreement or alter future comments' visibility rather than responding only to the comment at hand.
Focusing on these strategies---replying, deleting, and ignoring---therefore enables us to compare creator-controlled, comment-level responses. 
Table \ref{tab:moderation_actions} summarizes these three harm-addressing actions available to content creators along with their consequences on norm-violating comments, moderation intensity (i.e., how punitive the action is for the moderated commenter), and expected effort.

\begin{table}[ht]
\centering
\caption{A Comparison of Delete, Reply, and Ignore Responses to an Inappropriate Comment}
\resizebox{0.98\textwidth}{!}{
\begin{tabular}{l l l l l }
\hline
\textbf{Action} & \textbf{Consequences} & \textbf{Moderation Intensity} & \textbf{Expected Effort} \\
\hline
Delete & The comment becomes unavailable & High (Removal) & Medium \\
Reply & The comment and its reply are visible & Medium (Warning) & High \\
Ignore & The comment remains visible & Low (None) & Low \\
\hline
\end{tabular}
}
\label{tab:moderation_actions}
\end{table}

HCI scholars have often examined these moderation practices through the lens of safety, studying how effective they are in protecting users against online harms~\cite{zhang2023cleaning,gillespie2018custodians,kalch2017replying,jhaver2023douserswant,john2024classification}. 
However, it remains unclear how creators comparatively evaluate these practices and how their evaluation extends beyond safety-related goals.
Motivated by these questions, we examine two additional dimensions that may shape creators’ evaluations and adoption choices across moderation strategies: impression management benefits and algorithmic incentives. 
In the following sections, we elaborate on all three of these dimensions and contend why they are crucial for understanding how creators perceive and adopt different moderation strategies.

\subsection{Content Creators’ Perceptions of Online Safety}
Social media users commonly encounter harmful content and behaviors, such as hate speech, slurs, misinformation, and impersonation~\cite{thomas2021sok,citron2014addressing,scheuerman_framework_2021,samermit2023millions}. 
These experiences can produce a wide range of harms, including physical, emotional, relational, and financial harms~\cite{scheuerman_framework_2021}. 
Among these, emotional harms are particularly prevalent and may  manifest as stress, anxiety, or trauma, either immediately or over time through repeated exposure ~\cite{scheuerman_framework_2021,thomas2022s}. 
Moreover, these effects are not limited to direct targets but also extend to bystanders who are exposed to such interactions~\cite{scheuerman_framework_2021}. 
As such, emotional safety is a meaningful dimension of user protection and online safety that merits further examination in content moderation research.

Both platform- and community-level content moderation mechanisms can support users' emotional safety, most prominently by carefully removing harmful content~\cite{sasse2023breaking,murphy2026can}.
Individual users can also strive to achieve emotional safety through enacting personalized measures that accommodate their varying perceptions of harms and specific safety needs~\cite{scott2023trauma,rashed2026if,jhaver2023personalizing}.
Across these different modes, scholars have focused on securing emotional safety for vulnerable individuals, such as those dealing with mental health conditions (e.g., eating disorders, depression) or those at high risk of online harassment (e.g., women and BIPOC users) ~\cite{andalibi2017sensitive,blackwell2019harassment}.
Yet, existing work has primarily evaluated individual moderation measures based on their emotional safety outcomes and used these findings to inform platform safety designs~\cite{jhaver2018online,sasse2023breaking,thomas2021sok}.
This leaves a gap in understanding whether individuals anticipate these outcomes, or how they comparatively evaluate and choose among different strategies to achieve emotional safety.

This gap is especially important in the context of content creators, who are granted direct authority to manage harmful interactions within their own spaces.
Given that creators actively make moderation decisions and often express confidence in their ability to do so~\citep{samermit2023millions}, understanding how they perceive the emotional consequences of these actions is critical.
To address this gap, this study examines how different content creator–led moderation strategies (deleting, replying, and ignoring) are comparatively perceived on the dimension of emotional safety, and how these perceptions affect moderation adoption.
Prior literature has shown that removing hate speech makes it no longer visible and thus increases users' perceived safety in online communities~\cite{sasse2023breaking,gillett2022safety}.
More broadly, visibility-reduction measures are commonly used in content moderation to limit users’ exposure to harmful content and promote online safety~\cite{grimmelmann2015virtues}. 
Because deleting removes hateful comments from public view, whereas replying to or ignoring them leaves them visible, deleting provides a greater reduction in visibility than alternative strategies.
Therefore, we expect creators to perceive deleting as more beneficial for achieving emotional safety compared to replying or ignoring, leading to the following hypotheses:

\vspace{0.2cm}

\textit{Hypothesis 1a: Content creators perceive deleting hate speech comments as more beneficial for achieving emotional safety than replying to them.} 
\vspace{0.2cm}

\textit{Hypothesis 1b: Content creators perceive deleting hate speech comments as more beneficial for achieving emotional safety than ignoring them.} 
\vspace{0.2cm}

In comparing visibility-securing strategies, we expect that replying to hate speech comments (as opposed to ignoring them) clearly signals opposition~\cite{murumaa2021misogynist,im2022women} and may reduce the likelihood of further hate speech~\cite{bahador2021countering,obermaier2023ll,garland2022impact}.
This is in line with ~\citet{leets2002experiencing}, who found that active interventions against hate speech can mitigate emotional harm and enhance individuals’ sense of safety compared to remaining passive.
Given these demonstrated effects, we propose the following hypothesis:

\vspace{0.2cm}
\textit{Hypothesis 1c: Content creators perceive replying to hate speech comments as more beneficial for achieving emotional safety than ignoring them.}
\vspace{0.2cm}

\subsection{Content Creators' Impression Management}
Online users adjust their behaviors, such as interpersonal interactions and content sharing, in relation to the images they seek to project~\cite{chou2022content,rahmatulah2024impression,ellison2006managing,lang2015just,wohn2020audience}.
These behavioral adjustments are informed by users’ perceptions of imagined audiences~\cite{litt2012knock,litt2016imagined,marwick2011tweet}, based on their understanding of how their activities are interpreted by others~\cite{heung2024vulnerable}.
These audience-aware behavioral practices reflect impression management, a concept introduced by Goffman~\cite{goffman2002presentation}.

Impression management is particularly salient for content creators, as the images they project shape the extent and quality of audience engagement with their content~\cite{soneji2024feel,ma2023multi,wu2019agent}.
Such engagement is closely tied to monetary rewards~\cite{tafesse2023content,wohn2020audience}, making impression management economically consequential.
Accordingly, prior research has examined creators’ efforts to secure audience attention and generate revenue through impression management~\cite{abidin2016visibility,duffy2021nested,uttarapong2021harassment}. 
Scholars have found that creators make intentional choices about when and what content to upload based on their understanding of audience preferences~\cite{heung2024vulnerable,wohn2020audience}.
In particular, they maintain appealing self-presentations.
For instance, they adjust their tone, communication style, and interactions to align with their audience's demographics (e.g., age and gender)~\cite{wohn2020audience}.
They also tailor these behaviors based on their relationship with different audience members, ranging from family and supporters to lurkers and trolls ~\cite{heung2024vulnerable,olsson2022architectures}. 
Even when facing negative audience feedback, creators seek to project an authentic image by candidly responding to critical comments in order to maintain favorable impressions of themselves~\cite{rahmatulah2024impression}.

While this prior work demonstrates that impression management guided by anticipated audiences plays a central role in shaping creators’ content development and audience interaction~\cite{wohn2020audience,heung2024vulnerable,olsson2022architectures}, less attention has been paid to how creators’ moderation strategies are shaped by their impression management concerns.
Given that creators often consider impression management as a key factor in their decision-making, it is worth examining how different moderation strategies are perceived and adopted in terms of their benefits for managing the audience.
Accordingly, we examine impression management as a key dimension through which creators comparatively evaluate deletion, replying, and ignoring, and investigate how these perceptions shape their decisions to adopt each strategy.

Prior work has documented individuals' strategic use of verbal responses to repair or protect their public image when faced with accusations or criticism~\cite{benoit2014accounts}.
In particular, in an online context, replying to negative comments can provide creators with an opportunity to enact desired personas, such as by adopting a humorous tone~\cite{shim2025pin} or using sarcasm~\cite{ziegele2020not}.
Further, when hostile interactions occur, addressing hostility responses are more likely to be evaluated positively by fellow in-group members who strongly identify with the targeted group~\cite{kaiser2009group}.
This indicates that replying can serve as a tool for creators to achieve their impression management strategy.
In contrast, comment deletion may offer limited impression management benefits, as it may trigger backlash from audiences or controversy surrounding censorship and the suppression of speech, thereby raising free-speech concerns~\cite{juneja2020through,john2024classification,myers2018censored}.
Additionally, ignoring may help maintain a neutral or composed image by avoiding visible conflict, but it provides fewer opportunities for strategic persona crafting as compared to replying.
Accordingly, we propose the following hypothesis:

\vspace{0.2cm}

\textit{Hypothesis 2a: Content creators perceive replying to hate speech comments as promoting a more positive public image than deleting them.} 
\vspace{0.2cm}

\textit{Hypothesis 2b: Content creators perceive replying to hate speech comments as promoting a more positive public image than ignoring them.} 
\vspace{0.2cm}

Beyond creators’ strategic persona crafting, moderation strategies may also shape audience impressions by altering the visibility of harmful feedback.
Users’ understanding and interpretation of content are strongly shaped by accompanying comments~\cite{yeom2020meta_comments_effects}.
In some cases, other users’ feedback is weighted as more credible information than creators' posted content itself~\cite{deandrea2019influence,lane2022antecedents}, thereby influencing audience impressions of posts.
Given this influence, creators may strategically remove harmful comments to manage how both their content and their broader channel are perceived. Indeed, prior research has found that 47\% of social networking users report deleting comments to maintain a positive impression of their accounts~\cite{madden_smith_2010_reputation}.
Building on this, we argue that creators may view comment deletion as more beneficial for impression management than taking no action because it reduces the influence of harmful comments on audience perceptions.
By contrast, ignoring such comments may be publicly interpreted as indifference toward their comment sections, or as a lack of care in managing their public image.
Accordingly, we hypothesize:

\vspace{0.2cm}
\textit{Hypothesis 2c: Content creators perceive deleting hate speech comments as promoting a more positive public image than ignoring them.} 
\vspace{0.2cm}

\subsection{Content Creators' Visibility Being Shaped by Algorithmic Recommendation Systems}
Platform algorithms shape content engagements by determining which content reaches audiences and how widely it circulates~\cite{verwiebe2024algorithm,choi2023creator}.
These algorithmically-mediated engagements play a critical role in shaping content creators’ popularity and monetization opportunities on social media~\cite{wu2019agent,choi2023creator}.
To leverage algorithms effectively, creators actively engage in sense-making about how algorithms function, based on their own and others' experiences~\cite{choi2023creator,bishop2019managing,eslami2016first,ma2023multi,wu2019agent,gelman2011concepts}.
These understandings, in turn, influence creators' behavior and decision-making as they seek to improve their algorithmic visibility~\cite{french2017s,cotter2024practical,choi2023creator,wu2019agent,bertaglia2021clout,harper2021conspiracy,shim2025pin,scolere2018constructing}.

For instance, based on their experience, some creators perceive recommendation algorithms as favoring provocative or controversial content, and adjust their content creation practices accordingly~\cite{wu2019agent}.
Those who have this belief may strategically produce attention-grabbing but lower-quality content, such as conspiracy-oriented posts or clickbait videos, to maximize visibility~\cite{bertaglia2021clout,harper2021conspiracy}.
Beyond the topics of posted content, creators often believe that higher values on engagement metrics such as views, likes, comments, and shares lead algorithms to promote content more widely~\cite{glotfelter2019algorithmic}. 
In response, creators directly encourage users to reply to their posts to increase the number of comments~\cite{glotfelter2019algorithmic}.
They also use popular hashtags known to attract engagements (e.g., \#fyp and \#foryou on TikTok)~\cite{klug2021trick}. 
Beyond the volume of engagement, Shim and Jhaver~\citep{shim2025pin} found that some creators perceive toxic comments as potentially favored by recommendation algorithms. 
Creators who hold this belief may tolerate and retain inappropriate comments on their posts in pursuit of algorithmic benefits~\citep{shim2025pin}.

These strategic behaviors demonstrate that creators modify their content production and engagement practices in response to their perceptions of recommendation algorithms, with the aim of enhancing content visibility.
However, to our knowledge, limited research has examined how creators perceive different moderation strategies in relation to the algorithmic incentives they may provide. 
Building on the literature reviewed above, we hypothesize that creators' perception of the algorithmic incentives associated with different moderation strategies vary according to the extent to which each strategy preserves the visibility of negative comments and enables subsequent audience engagement.
In particular, as noted in Section \ref{relatedwork_2.1}, replying to negative comments may fuel further controversy, and generate additional engagement, compared to deleting or ignoring them.
We therefore hypothesize: 

\vspace{0.2cm}
\textit{Hypothesis 3a: Content creators perceive replying to hate speech comments as providing greater algorithmic visibility benefits than deleting them.}
\vspace{0.2cm}

\textit{Hypothesis 3b: Content creators perceive replying to hate speech comments as providing greater algorithmic visibility benefits than ignoring them.}
\vspace{0.2cm}

Moreover, deleting and ignoring differ in whether negative comments remain visible.
Deleting inherently reduces the number of visible comments in the comment thread, whereas ignoring preserves the presence of negative comments and the engagement signals associated with them.
Thus, creators may perceive ignoring as offering greater algorithmic benefits than deleting:

\vspace{0.2cm}
\textit{Hypothesis 3c: Content creators perceive ignoring hate speech comments as providing greater algorithmic visibility benefits than deleting them.}
\vspace{0.2cm}

\subsection{Creators' Likelihood of Adopting Comment Moderation Strategies}
All the hypotheses we have developed above specifically focus on how creators compare deleting, replying, and ignoring hate comments in terms of emotional safety, impression management, and algorithmic incentives (RQ1). 
RQ2 extends this inquiry by asking how these perceived benefits predict creators’ likelihood of adopting each strategy.

With regard to safety, prior research shows that moderation strategies can reduce exposure to harmful content and support users’ emotional well-being~\cite{sasse2023breaking,samermit2023millions,thomas2022s}; these latter outcomes may thus, in turn, influence moderation. 
Work on impression management similarly shows that creators modify their behaviors in response to public image concerns shaped by imagined audiences~\cite{abidin2016visibility,wohn2020audience,heung2024vulnerable,rahmatulah2024impression}, indicating that anticipated reputational consequences may also shape moderation adoption.
Research on algorithmic visibility further demonstrates that creators’ decisions are shaped by their understandings of recommendation systems, engagement metrics, and visibility incentives~\cite{cotter2024practical,choi2023creator,wu2019agent,glotfelter2019algorithmic,shim2025pin}, suggesting that anticipated algorithmic consequences may likewise influence their adoption of moderation strategies.

Taken together, this prior literature offers theoretical grounds for expecting all three dimensions to influence creators' willingness to adopt different moderation strategies. 
However, prior work has largely examined safety, impression management, and algorithmic visibility in isolation, offering limited insight into their relative and combined effects on moderation adoption. 
Therefore, rather than proposing directional hypotheses for RQ2, we conduct an open inquiry regarding how these theoretically motivated dimensions together predict creators’ adoption likelihood for each strategy.
\section{Methods}
\label{methods}
\subsection{Participant Recruitment}
We conducted a within-subjects online survey experiment, administered via Qualtrics, to answer our research questions.
Our participants were adults aged 18 and older, and self-identified as content creators who regularly produce content on social media sites.
The survey was distributed via Prolific\footnote{\url{https://www.prolific.com/}} and collected a total of 755 responses.
Among the 755 respondents, 51 did not proceed with the survey as they failed the screening criteria (self-identifying as active content creators, detailed below).
Additionally, we excluded 120 responses that failed attention checks, resulting in a total of 584 valid responses collected and analyzed for the main study.
In Appendix A, Table \ref{sec:appendix_survey_sample1} shows the demographic breakdown of our survey respondents, and Table \ref{sec:appendix_survey_sample2} further summarizes their content creation attributes, such as the platform they use, topics covered, and follower counts.
Participants who passed the screening question and completed the survey were paid \$1.5 and those who did not get past the screening question were paid \$0.14.
Rutgers University's
Institutional Review Board (IRB) reviewed the study and categorized its status as exempt on December 5, 2025.

\subsection{Survey Flow}
\label{surveyflow}
Our target population consisted of currently active content creators. To reach them, we used Prolific’s prescreening filters to recruit participants who identified as either ``Influencer'' or ``Video content creator''.
To further confirm participant eligibility, participants completed the following two screening questions after providing informed consent:
\begin{enumerate}
    \item I consider myself a content creator.\footnote{We adopted this question from prior literature, which conducted research on creator population~\cite{ma2023multi}.}
    \item I have created or shared content on social media or streaming platforms at least once in the past month (e.g., a post, reel, video, or live stream).
\end{enumerate}

Eligible participants then proceeded to the main survey, where we presented them with a hypothetical scenario of encountering hate comments designed to elicit their evaluation of different moderation strategies:

\textit{``Imagine that you post a submission on social media and subsequently receive around 100 comments on your post. You notice that about 10 of these comments, posted by people you do not know, are inappropriate. These comments do not contain swear words, but they subtly belittle aspects of your identity, such as your gender, race, religion, etc.''}

This scenario operationalizes hate speech as inappropriate comments that belittle aspects of creators' identities. 
We intentionally used a less extreme form of hate speech to increase the ecological validity of the study~\cite{kihlstrom2021ecological}, i.e., the extent to which an experimental setting or scenario resembles real-world situations.
Severe hate speech is often removed automatically at the platform-level, leaving creators with limited opportunity to make their own moderation decisions for that speech.
By presenting a scenario in which the comments remained plausible targets of creator intervention, we were able to examine creators' everyday moderation judgments and underlying decision-making heuristics.

After presenting this scenario, we asked participants a set of questions related to three strategic responses to the inappropriate comments therein: deleting, replying to, and ignoring.
The survey items measured creators’ perceptions of each strategy’s benefits across three dimensions: (a) emotional safety, (b) public image construction, and (c) content promotion through recommendation algorithms.
Additionally, we also assessed participants' likelihood of adopting each strategy.
Further, we collected demographic information on age, race, gender, education level and channel-related information, such as primary platform(s) used for content creation, number of followers, and content topic(s).

\subsection{Measures}
Participants were instructed to consider all three strategies (replying, deleting, and ignoring), which were presented in a randomized order. For each strategy, participants were required to evaluate the three factors (a)–(c) described in Section~\ref{surveyflow}. 
To measure each factor, we adopted and modified items from prior research when available and created new items informed by relevant literature when necessary.
To address RQ 2, we also assessed participants' likelihood of adopting each moderation strategy. 
Participants rated all items on a 7-point Likert scale (1 = strongly disagree / strongly unlikely; 7 = strongly agree / strongly likely).
These survey items are detailed below.

First, we conceptualized emotional safety as creators’ anticipated reduction in emotional harm resulting from the enactment of a given moderation strategy.
To measure this, we developed five measurement items adapted from \citet{scheuerman2021framework}, who identified nine dimensions of harm severity.
From these, we selected the five dimensions most relevant to creators' harm-addressing behaviors---intensity, agency, urgency, vulnerability, and sphere---and modified them to assess the extent to which each strategy was expected to reduce emotional harm.

Prompt: ``I expect that (\textit{replying to/deleting/ignoring}) these comments would''
\begin{enumerate}
\item reduce my emotional distress. (Intensity)
\item make me feel capable of managing my emotions. (Agency)
\item help me restore my emotional safety more quickly. (Urgency)
\item reduce the risk of further emotional harm. (Vulnerability)
\item have consequences for the safety of other users on the platform beyond myself. (Sphere)
\end{enumerate}

Prior work suggests that anticipating conflict escalation increases perceived safety risks and shapes individuals’ willingness to intervene~\cite{davidovic2023intervene}.
Because moderation actions may prolong or escalate a harmful interaction, we considered not only whether a strategy could reduce the immediate harm posed by a comment, but also whether it might worsen the situation.
We therefore added an item assessing the extent to which each strategy was expected to escalate the interaction:

\begin{enumerate}
\setcounter{enumi}{5}
    \item result in an escalation of the conflict. (Possible escalation)
\end{enumerate}

The six items demonstrated acceptable to good internal consistency across the three strategies (Cronbach's $\alpha$ value: deleting = 0.69, replying = 0.79, and ignoring = 0.70) and were averaged to create a composite emotional safety score for each strategy.

\hspace{0.5em}

To measure the second dimension, perceived public image benefits, we adopted and modified five measurement items out of 15 on the impression management scale developed by~\citet{bolino1999measuring}.
Although this original scale was developed to measure individuals' impression management behaviors in organizational contexts, it has been widely adopted across other domains, such as athlete branding~\cite{na2020exploring}, social media use~\cite{tuck2024social,rui2013strategic}, and entrepreneurship research~\cite{calic2023dark}.
Given the demonstrated versatility of this instrument, we deployed it for our use by adapting one item corresponding to each of the five most relevant impression management aspects---ingratiation, self‐promotion, exemplification, supplication, and intimidation---to measure how each moderation strategy was perceived to shape creators' public image.
The items were presented with the following prompt:

Prompt: ``I expect that (\textit{replying to/deleting/ignoring}) these comments would''

\begin{enumerate}
    \item help viewers see me as a nice person. (Ingratiation)
    \item help viewers see me as a competent creator. (Self-promotion)
    \item make viewers see me as morally exemplary. (Exemplification)
    \item help viewers feel sympathy for me. (Supplication)
    \item clearly signal that I do not tolerate inappropriate comments. (Intimidation)
\end{enumerate}

The five items demonstrated good internal consistency across the three strategies (Cronbach's $\alpha$ value: deleting = 0.81, replying = 0.84, and ignoring = 0.78) and were averaged to create a composite impression management score for each strategy.

\hspace{0.5em}

Third, to measure creators’ perceptions of how each moderation strategy influences the promotion of their content through recommendation algorithms, we developed new measurement items. 
We conceptualized recommendation algorithms as a system that distributes and prioritizes content to facilitate end-users' discovery of posts~\cite{NatlAcadScis_2024,narayanan2023understanding}.
Building on Sun et al.'s work on algorithmic visibility, we distinguished two dimensions of algorithmic impact: rank and time~\citep{sun2025dynamical}.
Specifically, \textit{rank} refers to whether content is promoted to a higher position in the feed, capturing algorithmic prioritization, whereas \textit{time} refers to how long or how repeatedly content remains promoted, capturing the duration of algorithmic visibility~\cite{sun2025dynamical}.

Beyond rank and time, algorithmic visibility also varies in its audience reach.
Prior work shows that creators strategically manage visibility both to maintain existing audiences and reach new ones~\citep{cotter2019playing}. 
This distinction maps onto two forms of algorithmic distribution: the ranking of content within existing followers' feeds and the recommendation of content to users who do not already follow the creators~\citep{NatlAcadScis_2024,narayanan2023understanding}.
Bringing these aspects together, we operationalized recommendation algorithm incentives by assessing perceived changes in rank and time separately for two audience groups: followers and non-followers.

Prompt: ``I expect that (\textit{replying to/deleting/ignoring}) these comments would:''

\begin{enumerate}
    \item increase the likelihood that recommendation algorithms prioritize my content for people who follow me. (Visibility: Followers' feed)
    \item increase how long recommendation algorithms promote my content to people who follow me. (Duration: Followers' feed)
    \item increase the likelihood that recommendation algorithms promote my content to people who do not currently follow me. (Visibility: Non-followers' feed)
    \item increase how long recommendation algorithms promote my content to people who do not currently follow me. (Duration: Non-followers' feed)
\end{enumerate}

The four items demonstrated good internal consistency across the three strategies (Cronbach's $\alpha$ value: deleting = 0.94, replying = 0.90, and ignoring = 0.94) and were averaged to create a composite algorithmic incentive score for each strategy.

\hspace{0.5em}

Beyond the measurement items assessing the perceived algorithmic influence of each strategy, we included items to examine creators’ beliefs about the algorithmic consequences of receiving inappropriate comments.
Because these broader beliefs may inform creators’ evaluations of different moderation strategies, we measured them separately from the strategy-specific items.
Measuring these beliefs helps contextualize how strategies such as ignoring inappropriate comments are interpreted by creators. 
For instance, if participants believe that inappropriate comments increase algorithmic visibility, leaving those comments visible may be understood as a strategic form of non-intervention that preserves them rather than simply indicating inaction. 

Participants responded to the following prompt: ``Compared to receiving no inappropriate comments, when your post receives 10 inappropriate comments out of 100, does this increase or decrease:'' 


\begin{enumerate}
    \item the likelihood that your content is promoted by recommendation algorithms to people who follow you?
    \item how long your content is promoted by recommendation algorithms to people who follow you?
    \item the likelihood that your content is promoted by recommendation algorithms to people who do not currently follow you?
    \item how long your content is promoted by recommendation algorithms to people who do not currently follow you?
\end{enumerate}

Participants rated these items on a 7-point Likert scale from Strongly decrease to Strongly increase.
These four items demonstrated good internal consistency (Cronbach's $\alpha$ = 0.88).

\hspace{0.5em}

Lastly, we asked the following questions to assess the likelihood of adopting and the cognitive burden of enacting each strategy.\vspace{0.5em}

\textit{Likelihood of adoption:}
\begin{enumerate}
    \item How likely are you to (reply to/delete/ignore) these comments?
\end{enumerate}
\vspace{0.5em}

\textit{Perceived burden:}
\begin{enumerate}
    \item How much effort would it take to (reply to/delete/ignore) these comments?
\end{enumerate}
\vspace{0.5em}



\section{Analysis and Findings}
We conducted repeated-measures ANOVA tests to address RQ1, which examines whether creators' perceptions of emotional safety, impression management benefits, and algorithmic incentives differed across deleting, replying, and ignoring strategies. 
To provide additional context for interpreting the results related to algorithmic incentives, we also analyzed participants' beliefs about whether the presence of negative comments increases or decreases the possibility of algorithms promoting their content.
 Section \ref{finding_RQ1} reports the results of these analyses. 

For RQ2, we employed regression models to examine how creators' perceptions of emotional safety, impression management, and algorithmic incentives tied to each moderation strategy were associated with the likelihood of adopting that strategy. 
In a separate analysis, we conducted a repeated-measures ANOVA test to compare the perceived effort required to enact the three moderation strategies.
The results of these analyses are presented in Section \ref{finding_RQ2}.

\subsection{Differences in Perceived Emotional Safety, Impression Management Benefit, and Algorithmic Advantage Across Moderation Strategies}
\label{finding_RQ1}
We first conducted repeated measures ANOVA tests to address RQ1. 
The results showed that participants' perceptions of emotional safety, impression management benefits, and algorithmic incentives were significantly different across the three moderation strategies.
Table ~\ref{table:ANOVA_Result} presents the mean rating for each strategy across the three dimensions, corresponding ANOVA results, and partial eta squared ($\eta_p^2$) values, which indicate the magnitude of the differences across strategies.
Further, Tukey-adjusted post-hoc comparisons showed the differences across pairs of strategies.

\begin{table}[htbp]
\centering
\caption{Mean and Standard Deviation of the Impact of Each Moderation Strategy (on a scale of 1-7) on Three Dimensions of Creator Concerns. Repeated-Measures ANOVA Results for Perceived Outcomes Across Moderation Strategies Are Also Shown.}
\label{table:ANOVA_Result}
\small
\setlength{\tabcolsep}{4pt}
\renewcommand{\arraystretch}{1.1}

\resizebox{0.85\textwidth}{!}{%
\begin{tabular}{lccccc}
\toprule
& \multicolumn{3}{c}{Mean (SD)} & \multicolumn{2}{c}{ANOVA results} \\
\cmidrule(lr){2-4}
\cmidrule(lr){5-6}
Dimension 
& Deleting
& Replying
& Ignoring
& \textbf{$F$} 
& \textbf{$\eta_p^2$} \\
\hline
Safety
& 5.03 (1.067) 
& 3.66 (1.246)  
& 4.65 (1.128) 
& $F(1.938, 1129.878) = 227.696$$^{***}$  
& .281 \\

Impression
& 4.50 (1.196)
& 4.40 (1.283)
& 4.20 (1.223)
& $F(1.886, 1099.426) = 10.715$$^{***}$  
& .018 \\

Algorithm
& 4.05 (1.513)
& 4.58 (1.344)
& 4.23 (1.456)
& $F(1.942, 1132.219) = 28.484$$^{***}$  
& .047 \\
\bottomrule
\end{tabular}%
}

\vspace{0.2em}
\begin{minipage}{0.85\linewidth}
\footnotesize
\textit{Note.} For all three outcomes, we report Greenhouse--Geisser-corrected results because the assumption of sphericity was violated.
$N = 584$, $\eta_p^2$ = partial eta squared.
$^{*}p < .05$, $^{**}p < .01$, $^{***}p < .001$.
\end{minipage}
\end{table}


For perceived emotional safety, \textit{deleting} was rated as most beneficial for achieving safety and \textit{replying} was rated as the least beneficial (deleting--replying: $M_{\mathrm{diff}} = 1.37$, $p < .001$; deleting--ignoring: $M_{\mathrm{diff}} = 0.37$, $p < .001$; replying--ignoring: $M_{\mathrm{diff}} = -1.00$, $p < .001$).
These findings support H1a and H1b, which predicted that deleting would be more beneficial for achieving safety than replying and ignoring respectively. 
However, H1c was not supported. 
Contrary to our prediction that replying would be more beneficial for achieving safety than ignoring, the results showed ignoring as being more beneficial for achieving emotional safety than replying.

For impression management benefits, the difference between \textit{replying} and \textit{deleting} was not significant (replying--deleting: $M_{\mathrm{diff}} = -0.10$, $p = .31$), but both \textit{replying} and \textit{deleting} were perceived as more beneficial than \textit{ignoring} (replying--ignoring:  $M_{\mathrm{diff}} = 0.20$, $p = .009$; deleting--ignoring: $M_{\mathrm{diff}} = 0.30$, $p < .001$).
These findings support H2b and H2c, which predicted that replying and deleting would be perceived as more beneficial than ignoring respectively.
However, H2a, which predicted that replying would be perceived as more beneficial than deleting was not supported.

For perceived algorithmic benefits, participants rated \textit{replying} as significantly higher than \textit{deleting} and \textit{ignoring}, and \textit{ignoring} as significantly higher than \textit{deleting} (replying--deleting: $M_{\mathrm{diff}} = 0.53$, $p < .001$; replying--ignoring: $M_{\mathrm{diff}} = 0.34$, $p < .001$; ignoring--deleting: $M_{\mathrm{diff}} = 0.18$, $p = .031$).
These findings support H3a, H3b and H3c.

Comparing effect sizes across the three dimensions, the differences were largest for perceived safety ($\eta_p^2$ = .281), suggesting that creators differentiated deleting, replying, and ignoring most strongly along the safety dimension.
Perceptions of impression management benefits and algorithmic incentives also differed significantly across strategies, but the effect sizes were comparatively smaller (impression $\eta_p^2$ = .018; algorithm $\eta_p^2$ = .047).

To sum up, these results show that participants evaluated the three strategies differently in terms of perceived safety advantages, impression management benefits, and algorithmic incentives; however, the clearest distinction among the strategies emerged in how they were perceived to offer safety.

\subsubsection{Creators' Belief in How Negative Comments Influence Recommendation Algorithm}
\label{finding_algorithm}
To examine whether participants associated the volume of negative comments with algorithmic advantages, we averaged 4 items that asked creators’ beliefs about the algorithmic consequences of inappropriate comments ($M = 4.58$, $SD = 1.19$),
and ran a one-sample $t$-test against the scale midpoint of 4 (neither an increase nor a decrease in algorithmic incentives).
The result was statistically significant ($t(583) = 11.81$, $p < .001$) with a moderate effect size (Cohen's $d = .49$).
Thus, participants ratings were significantly above the neutral midpoint, indicating a tendency to perceive a higher volume of negative comments as algorithmically beneficial, although this tendency was modest in magnitude.

\subsection{Influence of Each Dimension of Creator Concerns on Moderation Adoption}
\label{finding_RQ2}

To address RQ2, we conducted regression analyses. 
We included participants' gender\footnote{Gender was measured using four categories: woman, man, non-binary, and prefer not to answer~\cite{spiel2019how}.} and follower size\footnote{Follower size was coded into three categories: nano creators (10,000 or fewer followers), micro creators (10,001--100,000 followers), and macro creators (more than 100,000 followers). This categorization was based on~\citet{campbell2020more}.} as control variables, since prior research suggests that creators' gender and number of followers may influence their adoption of moderation tools~\cite{schoenebeck2021drawing, shim2025pin}.
We also included the perceived effort of enacting each moderation strategy as a control variable, given that the likelihood of adopting any information technology is often impacted by the perceived effort required~\cite{davis1989perceived}.

Table~\ref{RQ2_delete} shows the regression results on how perceived safety, impression management benefits, and algorithmic incentives are associated with the likelihood of deleting negative comments.
We found that perceived safety had the strongest association with adoption likelihood ($\beta = .351$, $p < .001$), followed by impression management ($\beta = .201$, $p < .001$).
However, algorithmic incentives were not significantly associated with deletion adoption ($\beta = .001$, $p = .977$).

\begin{table}[htbp]
\centering
\renewcommand{\arraystretch}{1}
\caption{Regression Results for the Effects of Perceived Dimensions on Deletion Strategy Adoption Likelihood}
\label{RQ2_delete}

\resizebox{0.85\textwidth}{!}{%
\begin{tabular}{ p{2cm} p{5cm} p{3cm} p{2cm}
}
\toprule
Strategy & Variable & $B$ (SE) & $\beta$ \\
\midrule

\multirow{12}{*}{Deleting}
& \multicolumn{3}{l}{\textit{Control variables}} \\
& \hspace{1em}Effort 
& .112 (.038)$^{**}$ 
& .108$^{**}$ \\

& \hspace{1em}Gender: Male
& -.257 (.149) 
& -.063 \\

& \hspace{1em}Gender: Non-binary 
& -.394 (.724) 
& -.020 \\

& \hspace{1em}Gender: Prefer not to answer
& .076 (1.051) 
& .003 \\

& \hspace{1em}Follower: Micro
& -.500 (.166)$^{**}$ 
& -.111$^{**}$ \\

& \hspace{1em}Follower: Macro
& -.200 (.369) 
& -.020 \\

& \hspace{1em}Follower: Prefer not to answer
& .307 (.817) 
& .014 \\

\cmidrule(lr){2-4}
& \multicolumn{3}{l}{\textit{Dimensions}} \\
& \hspace{1em}Safety 
& .661 (.079)$^{***}$ 
& .351$^{***}$ \\
& \hspace{1em}Impression 
& .338 (.077)$^{***}$ 
& .201$^{***}$ \\
& \hspace{1em}Algorithm 
& .002 (.055) 
& .001 \\

\midrule

Model fit &  

 Adjusted $R^2 = .244$ \\

\bottomrule
\end{tabular}
}
\vspace{2mm}
\begin{minipage}{0.8\linewidth}
\footnotesize
\textit{Note.} 
$B$ values are unstandardized coefficients, with standard errors in parentheses; $\beta$ values are standardized coefficients. Women and nano creators are the reference categories for gender and follower size, respectively.
$^{*}p < .05$, $^{**}p < .01$, $^{***}p < .001$.
\end{minipage}
\end{table}

Turning to replying, impression management showed the strongest association with adoption likelihood ($\beta = .248$, $p < .001$), followed by perceived safety ($\beta = .171$, $p < .001$). 
However, algorithmic incentives were not significantly associated with replying adoption ($\beta = .075$, $p = .085$) (see Table ~\ref{RQ2_Reply}).

\begin{table}[htbp]
\centering
\renewcommand{\arraystretch}{1}
\caption{Regression Results for the Effects of Perceived Dimensions on Replying Strategy Adoption Likelihood}
\label{RQ2_Reply}

\resizebox{0.8\textwidth}{!}{%
\begin{tabular}{ p{2cm} p{5cm} p{3cm} p{2cm} 
}
\toprule
Strategy & Variable & $B$ (SE) & $\beta$ \\
\midrule

\multirow{12}{*}{Replying}
& \multicolumn{3}{l}{\textit{Control variables}} \\
& \hspace{1em}Effort 
& .129 (.041) 
& .119$^{**}$ \\

& \hspace{1em}Gender: Male 
& -.138 (.153) 
& -.034 \\

& \hspace{1em}Gender: Non-binary 
& .536 (.741) 
& .027 \\

& \hspace{1em}Gender: Prefer not to answer
& 1.169 (1.078) 
& .042 \\

& \hspace{1em}Follower: Micro
& .032 (.170) 
& .007 \\

& \hspace{1em}Follower: Macro
& -.126 (.379) 
& -.013 \\

& \hspace{1em}Follower: Prefer not to answer
& -.977 (.836) 
& -.045 \\

\cmidrule(lr){2-4}
& \multicolumn{3}{l}{\textit{Dimensions}} \\
& \hspace{1em}Safety 
& .287 (.076)$^{***}$ 
& .180$^{***}$ \\
& \hspace{1em}Impression 
& .373 (.079)$^{***}$ 
& .241$^{***}$ \\
& \hspace{1em}Algorithm 
& .110 (.064) 
& .075 \\
\midrule

Model fit &  

Adjusted $R^2 = .183$ \\

\bottomrule
\end{tabular}
}
\vspace{2mm}
\begin{minipage}{0.85\linewidth}
\footnotesize
\textit{Note.} $B$ values are unstandardized coefficients, with standard errors in parentheses; $\beta$ values are standardized coefficients. Women and nano creators are the reference categories for gender and follower size, respectively.
$^{*}p < .05$, $^{**}p < .01$, $^{***}p < .001$.
\end{minipage}
\end{table}

For ignoring, perceived safety was the strongest predictor ($\beta = .301$, $p < .001$), followed by impression management benefits ($\beta = .181$, $p < .001$).
Here again, algorithmic incentives were not significantly associated with strategy adoption ($\beta = .004$, $p = .925$; see Table ~\ref{RQ2_Ignore}).

\begin{table}[htbp]
\centering
\renewcommand{\arraystretch}{1}
\caption{Regression Results for the Effects of Perceived Dimensions on Ignoring Strategy Adoption Likelihood}
\label{RQ2_Ignore}
\label{table:adoption_regression}

\resizebox{0.85\textwidth}{!}{%
\begin{tabular}{ p{2cm} p{5cm} p{3cm} p{2cm} 
}
\toprule
Strategy & Variable & $B$ (SE) & $\beta$ \\
\midrule

\multirow{12}{*}{Ignoring}
& \multicolumn{3}{l}{\textit{Control variables}} \\
& \hspace{1em}Effort 
& -.029 (.036) 
& -.031 \\

& \hspace{1em}Gender: Male 
& -.253 (.154) 
& -.063 \\

& \hspace{1em}Gender: Non-binary 
& .864 (.741) 
& .044 \\

& \hspace{1em}Gender: Prefer not to answer
& -.159 (1.078) 
& -.006 \\

& \hspace{1em}Follower: Micro
& .091 (.169) 
& .021 \\

& \hspace{1em}Follower: Macro
& .427 (.380) 
& .043 \\

& \hspace{1em}Follower: Prefer not to answer
& -1.444 (.840) 
& -.067 \\

\cmidrule(lr){2-4}
& \multicolumn{3}{l}{\textit{Dimensions}} \\
& \hspace{1em}Safety 
& .503 (.077)$^{***}$ 
& .286$^{***}$ \\
& \hspace{1em}Impression 
& .318 (.079)$^{***}$ 
& .196$^{***}$ \\
& \hspace{1em}Algorithm 
& .006 (.060) 
& .004 \\

\midrule

Model fit &  

Adjusted $R^2 = .182$ \\

\bottomrule
\end{tabular}
}
\vspace{2mm}
\begin{minipage}{0.8\linewidth}
\footnotesize
\textit{Note.} $B$ values are unstandardized coefficients, with standard errors in parentheses; $\beta$ values are standardized coefficients. Women and nano creators are the reference categories for gender and follower size, respectively.
$^{*}p < .05$, $^{**}p < .01$, $^{***}p < .001$.
\end{minipage}
\end{table}

Taken together, these findings suggest that creators' decisions to adopt moderation strategies were primarily associated with perceived emotional safety and impression management benefits, rather than with expectations of algorithmic advantages.

\begin{table}[htpb]
\centering
\caption{Repeated-Measures ANOVA Results for Perceived Effort Across Moderation Strategies}
\label{tab:effort_anova}
\renewcommand{\arraystretch}{1.1}
\begin{tabular}{lccccc}
\toprule
 & \multicolumn{3}{c}{Mean (SD)}  &  \multicolumn{2}{c}{ANOVA results}  \\
\cmidrule(lr){2-4}
\cmidrule(lr){5-6}
Dimension & deleting & Replying & Ignoring & $F$ & $\eta_p^2$ \\
\hline
Effort 
& 3.21 (1.94) 
& 4.70 (1.84) 
& 3.43 (2.13) 
& $F(1.92, 1119.68) = 96.37$$^{***}$  
& .142 \\
\bottomrule
\end{tabular}
\vspace{2mm}
\begin{minipage}{0.85\linewidth}
\footnotesize
\textit{Note.} Greenhouse--Geisser corrected result is reported because the assumption of sphericity was violated. 
$\eta_p^2$ = partial eta squared. $N = 584$.
$^{*}p < .05$, $^{**}p < .01$, $^{***}p < .001$.
\end{minipage}
\end{table}

\subsubsection{Creators' Perceived Effort to Enact Each Moderation Strategy}
We separately measured how much effort creators perceived each moderation strategy to require. 
Using these responses, we conducted a repeated-measures ANOVA to test whether perceived effort differed across the three strategies.

Our findings show that participants perceived the strategies as requiring significantly different levels of effort ($F(1.921, 1119.684) = 96.367$, $p < .001$, $\eta_p^2 = .142$) (See Table \ref{tab:effort_anova}).
Mauchly's test indicated that the assumption of sphericity was violated ($W = .959$, $p < .001$); thus, Greenhouse-Geisser corrections were applied to evaluate the ANOVA main effects. 
Subsequent Tukey-corrected pairwise comparisons showed that replying was perceived as significantly more effortful than deleting (replying--deleting: $M_{\text{diff}} = 1.49$, $p < .001$) and ignoring (replying--ignoring: $M_{\text{diff}} = 1.26$, $p < .001$). 
However, ignoring and deleting did not significantly differ from each other in perceived effort required (ignoring--deleting: $M_{\text{diff}} = .23$,  $p = .135$).

\section{Discussion}
\subsection{Fostering Creators to Become Responsible Stewards of Comment Spaces}
\label{discussion_middle-level}
Platforms delegate localized moderation authority to creators, such as the ability to delete comments, enabling them to respond to their heightened exposure to online harm~\cite{thomas2022s,matthews2025supporting}.
Such authority allows those directly exposed to harmful interactions to intervene and mitigate their effects. 
Given the scale of harmful online interactions, and the considerable interpretive labor required to assess them~\cite{roberts2014behind}, enabling creators to make timely moderation decisions within their own comment threads may help limit escalation and support safer online spaces. 
This aligns with our findings regarding how safety considerations shape creators' moderation perceptions and enactions (Sections \ref{finding_RQ1} and \ref{finding_RQ2}).
Moreover, because creators’ responses to online harm can influence broader audiences~\cite{chatzakou2017measuring,pater2016characterizations,thomas2021sok,scheuerman2021framework,samermit2023millions}, their moderation decisions may also contribute to shaping healthier online communities.

Our findings suggest that creators understand the exercise of moderation authority, at least in part, as a public-facing responsibility. 
Creators were more likely to adopt harm addressing strategies such as deleting, replying and ignoring, when they expected those actions to help them maintain a positive public image (Section~\ref{finding_RQ2}), suggesting that their responses to harmful comments are shaped by an awareness of how audiences may interpret them. 
This finding extends prior work showing that audience management permeates several of creators' everyday activities~\cite{rahmatulah2024impression,ellison2006managing,smallridge2016understanding,proudfoot2018saving} by demonstrating that it also shapes their decisions about whether and how to intervene in harmful interactions.

Specifically, our findings clarify what creators perceive as beneficial for managing their public image: deleting and replying were perceived as more beneficial for impression management than ignoring hate comments (Section~\ref{finding_RQ1}). 
Although deleting and replying strategies differ in form, they both allow creators to reshape audiences’ encounters with harmful comments: deleting reduces the number of hate comments to which audiences are exposed to, whereas replying makes creators’ counter-position visible and clarifies their role in contesting such speech.
Thus, for creators, a positive image is tied to being seen as a responsible steward of the comment space---someone who can recognize harm and maintain local norms by managing the presence, removal, or interpretation of harmful comments.

Taken together, our findings suggest that creators associate a positive public image, at least in part, with being perceived as responsible stewards of their comment spaces, and that this perceived responsibility influences their moderation choices.
Platforms should therefore attend to creators' perceptions of this role and design moderation tools that support their ability to act as responsible middle-level governance actors and have that role recognized by audiences~\cite{jhaver2023decentralizing}. 
It would also be useful to examine whether moderation strategies beyond deleting, replying, and ignoring, e.g., reporting and account muting, strengthen or undermine creators’ perception of themselves as responsible stewards, and exploring how those perception shape strategy adoption.

Moreover, creators are digital laborers who occupy an intermediary position between platforms and end-users, and continually adapt their practices to audience expectations~\cite{kang2025understanding,olsson2022architectures}.
Platforms should therefore maintain sustained communication with creators to understand how they currently interpret audience expectations when determining what counts as responsible moderation.
Such communication could help platforms identify how existing and emerging affordances affect creators’ willingness and ability to address harm without damaging their public image.

At the same time, creators’ concern with public impressions may also shape moderation in undesirable ways.
When moderation decisions are shaped by how audiences evaluate their actions, creators may selectively prioritize responses and address comments that are more visible, publicly legible, or likely to reinforce a favorable image, rather than those that most directly advance community safety.
This may leave less visible harms, marginalized users’ concerns, or interventions that carry reputational risks comparatively under-addressed.
Platforms should therefore avoid treating creator discretion as a substitute for clear governance principles. 
Because individuals differ in how they define harm and assess the emotional risks of responding to it~\citep{jhaver2023personalizing}, creator-led moderation tools should be grounded in clearly articulated platform norms and safety-centered policies while still allowing creators to adapt responses to local channel contexts.

\subsection{Deletion as an Emotional Safety Oriented Moderation Strategy}
Platforms often seek to foster safer online environments by removing or reducing the visibility of norm-violating content~\cite{grimmelmann2015virtues}.
Such intervention can be effective at the community level, e.g., prior research has found that removing hate speech can increase community members’ sense of safety~\cite{sasse2023breaking}. 
Extending this prior work, our study shows that the safety benefits of content removal also operate at the individual level: creators who were directly exposed and have authority to delete hate comments perceived deletion as providing greater emotional safety than ignoring or replying 
(Section \ref{finding_RQ1}).
Moreover, perceived emotional safety was the strongest predictor of creators' likelihood of adopting deletion, exceeding the effects of impression management and algorithmic concerns (Section \ref{finding_RQ2}).
This shows that creators' willingness to delete harmful comments strongly depends on whether they believe deletion will provide meaningful and sustained protection.

Together, these findings suggest that deletion functions as a particularly safety-oriented form of creator-led moderation. 
Platforms should therefore design deletion to fulfill its protective promise. 
Currently, although deletion reduces the immediate visibility of harmful content, it may also provoke backlash or retaliation~\cite{juneja2020through,john2024classification,myers2018censored}, potentially undermining creators’ perceptions of deletion as a safe and protective moderation strategy.
Platforms could mitigate these risks by connecting deletion with optional follow-up protections, such as restricting repeat offenders, placing similar comments in a review queue, or enabling creators to apply filters based on previously removed content.
By reducing repeated exposure and limiting opportunities for retaliation, such designs could strengthen creators’ confidence that deletion will genuinely improve their safety and, in turn, make them more willing to adopt it when needed.
These efforts may be especially helpful for creators who face disproportionate safety risks, e.g., LGBTQ+ and BIPOC creators~\cite{mariotto2026understanding,keum2024racial}.
However, to avoid over-moderation and prevent triggering further retaliation, such measures should remain reversible, governed by clear platform rules, and fully under creators’ control. 

\subsection{Ignoring as a Strategic Action to Address Online Hate Speech}
We found that creators perceived ignoring hate comments as more beneficial for achieving safety than replying to them.
Although creators perceived ignoring as less beneficial for impression management than deleting or replying, they still viewed it as having some positive value (Section \ref{finding_RQ1}).
These two dimensions---safety and impression management---also affect their decision to ignore hate comments, with safety showing the stronger association with adopting \textit{ignoring} (Section \ref{finding_RQ2}).
These findings show that ignoring may function as strategic non-engagement rather than merely passive inaction or negligence.

Although ignoring problematic content has been understood primarily as a way to avoid the social costs of intervention~\citep{tandoc2020diffusion}, our findings suggest that, for content creators, ignoring hate speech may also help achieve emotional safety and yield modest impression management benefits.
One possible explanation is that, by not responding to harmful comments, creators may refuse to reward hateful comments with additional attention~\cite{kiesler2012regulating}, thereby preventing further escalation. 
Similarly, ignoring may reflect creators' expectations that non-intervention helps them avoid appearing as overly controlling or engaging in  censorship~\cite{juneja2020through,john2024classification,myers2018censored}.

These findings suggest that platforms should not interpret harmful comment that remains visible as unmoderated, as creators may have made a strategic decision not to engage with it.
At the same time, because ignoring may otherwise be interpreted as indifference, oversight, or having limited capacity to respond~\cite{kiesler2012regulating}, platforms may benefit from recognizing that non-response can sometimes reflect a deliberate moderation choice. 
Because creators evaluated ignoring as providing greater benefits in safety than replying but less safety than deletion, the continued visibility of harmful comments may still raise safety concerns, suggesting a need for intermediate moderation options between ignoring and deletion. 
Thus, platforms could hide such comments from creators' own view, exclude them from recommended comments, or lower their ranking.
Such tools would preserve the non-engagement benefits of ignoring while incorporating some of the visibility-reduction benefits associated with deletion.

\subsection{Replying as a Communicative and Norm-Signaling Moderation Mechanism}
\label{discussion_reply}

Our results reveal that replying was perceived as providing less emotional safety than deleting or ignoring, but as offering impression-management benefits comparable to deletion and greater than ignoring.
Additionally, creators’ adoption of replying was more strongly associated with perceived impression management benefits than with perceived safety benefits (Section~\ref{finding_RQ2}).
These findings reveal a tradeoff: creator may adopt replying because it allows them to publicly signal their stance and manage audience impressions, even though they perceive it as less emotionally safe than deletion or ignoring.
Given that counterspeech may fuel further hostility or escalation because of its uncalibrated nature~\cite{shim2025pin,ping2024behind}, creators' lower safety evaluations of replying are understandable.
At the same time, its impression management benefits were comparable to those of deletion, suggesting that creators value replying as a visible means of signaling opposition and enforcing local norms.

Another notable insight here is that the expressive flexibility of replying did not produce greater perceived impression management benefits than deleting.
In other words, although creators can use humor, sarcasm, or supportive language to present themselves as caring toward their audiences, entertaining, or strong and thick-skinned~\cite{shim2025pin,ziegele2020not,cote2017can}, this expressive flexibility did not provide an additional impression-management advantage over removing the comment (Section~\ref{finding_RQ1}).
One possible interpretation is that creators’ perceived impression-management benefits of moderation may lie less in the expressive richness of a response than in its ability to make their governance role visible.
Both deletion and replying allow creators to demonstrate stewardship over their comment spaces, although they do so in different ways. 
This shared governance function may explain why the two strategies received comparable evaluations of their impression-management benefits.

This interpretation points to opportunities for low-engagement, high-legibility moderation tools that allow creators to make their stewardship visible without requiring sustained interaction with harmful commenters. 
Platforms could, for example, provide customizable norm signals through which creators indicate that a comment violates community expectations without composing a full reply or initiating a continuing exchange. 
This could serve as a warning stage before deletion, allowing creators to indicate that a comment violates community expectations and that continued violations may lead to removal or restriction.
Such features could preserve the public signaling benefits of replying while helping creators prevent replying from escalating into prolonged harmful interactions.

\subsection{Perceived Effort Is Not a Prominent Barrier to Moderation Adoption}
According to the Technology Adoption Model (TAM) ~\cite{davis1989perceived}, users' perceived effort is often treated as a barrier to adopting technology feature. 
Thus, we asked creators how much effort they perceive is required to implement each strategy and included this as a control variable in our analysis of adoption likelihood.
However, in our study, perceived effort was not negatively associated with creators' adoption of harm addressing strategies.
Instead, effort was positively associated with adopting deleting and replying strategies for hate speech comments, and was not significantly associated with adopting ignoring (Section \ref{finding_RQ1}).
This finding suggests that creators' moderation decisions cannot be understood simply as feature use driven by ease.
Instead, creators may be willing to bear the additional burden of moderation when they see such actions as meaningful for protecting themselves, enforcing boundaries, or communicating their governance role to audiences.
This interpretation aligns with prior work showing that creators are not driven just by efficiency-oriented moderation logics, but also seek to exercise moderation authority in ways that align with their values and contextual judgments~\citep{jhaver2022designing}.

Thus, we recommend that platforms distinguish between meaningful governance labor and unnecessary burdens, rather than treating support for creator moderation solely as a matter of minimizing effort or making moderation frictionless.
Since creators may be willing to invest effort in actions that help them maintain community norms, platforms should identify which forms of effort meaningfully support safety, boundary-setting, or public norm signaling, and which forms are redundant, emotionally taxing, or avoidable.

\subsection{Creators Have Nuanced Mental Models of Negative Comments and Recommendation Algorithms} 
Our findings show that creators generally associated a higher volume of negative comments with greater algorithmic benefits, expecting posts with more negative comments to receive higher visibility than those with no negative comments (Section \ref{finding_algorithm}). 
Their evaluations of moderation strategies reflected this broader belief: replying and ignoring, both of which preserve the visibility of negative comments, were perceived as more algorithmically advantageous than deleting 
(Section \ref{finding_RQ1}).
These findings broaden existing accounts of creators’ understandings of algorithmic visibility~\cite{glotfelter2019algorithmic,shim2025pin,klug2021trick} by showing that such beliefs extend beyond creators’ own content-production practices to include the volume and continued visibility of harmful audience engagement.

At the same time, although deleting was perceived as less algorithmically beneficial than replying and ignoring, its mean rating was close to the neutral midpoint of the scale.
Creators, therefore, may not view deletion as clearly harmful to visibility, even if they perceive strategies that preserve negative comments as relatively more advantageous.
This pattern suggests that creators' understanding of algorithmic incentives is not reducible to a simple engagement-based logic, in which more negative comments directly increase visibility and removing them necessarily decreases it.

Additionally, perceived algorithmic incentives were not significantly associated with creators' adoption of any moderation strategy (Section \ref{finding_RQ2}).
This finding adds nuance to prior work showing that social media platforms' algorithmic visibility logic shapes creators’ everyday practices~\cite{bertaglia2021clout,harper2021conspiracy,abidin2016visibility,duffy2021nested,uttarapong2021harassment}.
Thus, creators’ moderation practices should be examined as a distinct domain of actions---those pertinent to their responsibilities as middle-level governance actors---rather than treated as similar to the everyday visibility-seeking activities documented in prior work.

\subsection{Limitations and Future Work}
Our study focused on a hate speech scenario and examined only three moderation strategies: deleting, replying, and ignoring. 
This leaves room to explore how creators may evaluate a broader range of scenarios and strategies.
In particular, prior literature shows that efforts to increase engagement through algorithmic visibility shape creators’ behaviors~\cite{hodl2023content,lee2025managing,devito2022transfeminine} and suggests that algorithmic incentives may drive the adoption of social shaming-based moderation strategies, such as screenshotting~\cite{corry2021screenshot} and pinning negative comments~\cite{shim2025pin}.
Informed by these studies, future work should examine various moderation strategies, including social shaming practices, to identify the conditions under which algorithmic incentives play a more significant role in strategy adoption.


\begin{acks}
Awaiting paper acceptance.
\end{acks}

\bibliographystyle{ACM-Reference-Format}
\bibliography{references}

@article{kalch2017replying,
  title={Replying, disliking, flagging: How users engage with uncivil and impolite comments on news sites},
  author={Kalch, Anja and Naab, Teresa K},
  year={2017}
}

@article{atreja2023remove,
  title={Remove, reduce, inform: what actions do people want Social Media platforms to take on potentially misleading content?},
  author={Atreja, Shubham and Hemphill, Libby and Resnick, Paul},
  journal={Proceedings of the ACM on Human-Computer Interaction},
  volume={7},
  number={CSCW2},
  pages={1--33},
  year={2023},
  publisher={ACM New York, NY, USA}
}

@article{jhaver2023personalizing,
  title={Personalizing content moderation on social media: User perspectives on moderation choices, interface design, and labor},
  author={Jhaver, Shagun and Zhang, Alice Qian and Chen, Quan Ze and Natarajan, Nikhila and Wang, Ruotong and Zhang, Amy X},
  journal={Proceedings of the ACM on Human-Computer Interaction},
  volume={7},
  number={CSCW2},
  pages={1--33},
  year={2023},
  publisher={ACM New York, NY, USA}
}

@article{crawford2016flag,
  title={What is a flag for? Social media reporting tools and the vocabulary of complaint},
  author={Crawford, Kate and Gillespie, Tarleton},
  journal={New Media \& Society},
  volume={18},
  number={3},
  pages={410--428},
  year={2016},
  publisher={Sage Publications Sage UK: London, England}
}

@article{zhang2023cleaning,
  title={Cleaning up the streets: Understanding motivations, mental models, and concerns of users flagging social media posts},
  author={Zhang, Alice Qian and Montague, Kaitlin and Jhaver, Shagun},
  journal={arXiv preprint arXiv:2309.06688},
  year={2023}
}

@article{goldman2021content,
  title={Content moderation remedies},
  author={Goldman, Eric},
  journal={Mich. Tech. L. Rev.},
  volume={28},
  pages={1},
  year={2021},
  publisher={HeinOnline}
}

@article{jhaver2018online,
  title={Online harassment and content moderation: The case of blocklists},
  author={Jhaver, Shagun and Ghoshal, Sucheta and Bruckman, Amy and Gilbert, Eric},
  journal={ACM Transactions on Computer-Human Interaction (TOCHI)},
  volume={25},
  number={2},
  pages={1--33},
  year={2018},
  publisher={ACM New York, NY, USA}
}

@article{jhaver2023douserswant,
  title={Do users want platform moderation or individual control? Examining the role of third-person effects and free speech support in shaping moderation preferences},
  author={Jhaver, Shagun and Zhang, Amy X},
  journal={New Media \& Society},
  pages={14614448231217993},
  year={2023},
  publisher={SAGE Publications Sage UK: London, England}
}

@article{myers2018censored,
  title={Censored, suspended, shadowbanned: User interpretations of content moderation on social media platforms},
  author={Myers West, Sarah},
  journal={New Media \& Society},
  volume={20},
  number={11},
  pages={4366--4383},
  year={2018},
  publisher={SAGE Publications Sage UK: London, England}
}

@book{gillespie2018custodians,
  title={Custodians of the Internet: Platforms, content moderation, and the hidden decisions that shape social media},
  author={Gillespie, Tarleton},
  year={2018},
  publisher={Yale University Press}
}

@inproceedings{jhaver2022designing,
  title={Designing word filter tools for creator-led comment moderation},
  author={Jhaver, Shagun and Chen, Quan Ze and Knauss, Detlef and Zhang, Amy X},
  booktitle={Proceedings of the 2022 CHI conference on human factors in computing systems},
  pages={1--21},
  year={2022}
}

@article{geiger2016bot,
  title={Bot-based collective blocklists in Twitter: the counterpublic moderation of harassment in a networked public space},
  author={Geiger, R Stuart},
  journal={Information, Communication \& Society},
  volume={19},
  number={6},
  pages={787--803},
  year={2016},
  publisher={Taylor \& Francis}
}

@article{john2024classification,
  title={A Classification of Features for Interpersonal Disconnectivity in Digital Media: Block, Unfriend, Unfollow, Mute, Withhold, and Eject},
  author={Nicholas John},
  journal={Media and Communication},
  volume={12},
  number={Article 8716},
  year={2024},
  publisher={Cogitatio}
}

@article{corry2021screenshot,
  title={Screenshot, save, share, shame: Making sense of new media through screenshots and public shame},
  author={Corry, Frances},
  journal={First Monday},
  year={2021}
}

@article{schoenebeck2021drawing,
  title={Drawing from justice theories to support targets of online harassment},
  author={Schoenebeck, Sarita and Haimson, Oliver L and Nakamura, Lisa},
  journal={new media \& society},
  volume={23},
  number={5},
  pages={1278--1300},
  year={2021},
  publisher={Sage Publications Sage UK: London, England}
}

@inproceedings{thomas2022s,
  title={“It’s common and a part of being a content creator”: Understanding How Creators Experience and Cope with Hate and Harassment Online},
  author={Thomas, Kurt and Kelley, Patrick Gage and Consolvo, Sunny and Samermit, Patrawat and Bursztein, Elie},
  booktitle={Proceedings of the 2022 CHI conference on human factors in computing systems},
  pages={1--15},
  year={2022}
}

@article{smallridge2016understanding,
  title={Understanding cyber-vigilantism: A conceptual framework.},
  author={Smallridge, Joshua and Wagner, Philip and Crowl, Justin N},
  journal={Journal of Theoretical \& Philosophical Criminology},
  volume={8},
  number={1},
  year={2016}
}

@inproceedings{cai2024content,
  title={Content Moderation Justice and Fairness on Social Media: Comparisons Across Different Contexts and Platforms},
  author={Cai, Jie and Patel, Aashka and Naderi, Azadeh and Wohn, Donghee Yvette},
  booktitle={Extended Abstracts of the CHI Conference on Human Factors in Computing Systems},
  pages={1--9},
  year={2024}
}

@article{murumaa2021misogynist,
  title={Misogynist content expos{\'e} pages on Instagram: Five types of shamings, moderators and audience members},
  author={Murumaa-Mengel, Maria and Muuli, Liisi Maria},
  journal={Participations: Journal of Audience and Reception Studies},
  volume={18},
  number={2},
  pages={100--123},
  year={2021}
}

@article{french2017s,
  title={What's the folk theory? Reasoning about cyber-social systems},
  author={French, Megan and Hancock, Jeff},
  journal={Reasoning About Cyber-Social Systems (February 2, 2017)},
  year={2017}
}

@article{gelman2011concepts,
  title={Concepts and folk theories},
  author={Gelman, Susan A and Legare, Cristine H},
  journal={Annual review of anthropology},
  volume={40},
  number={1},
  pages={379--398},
  year={2011},
  publisher={Annual Reviews}
}

@article{marwick2011tweet,
  title={I tweet honestly, I tweet passionately: Twitter users, context collapse, and the imagined audience},
  author={Marwick, Alice E and Boyd, Danah},
  journal={New media \& society},
  volume={13},
  number={1},
  pages={114--133},
  year={2011},
  publisher={Sage Publications Sage UK: London, England}
}

@article{rashidi2020s,
  title={" It's easier than causing confrontation": sanctioning strategies to maintain social norms and privacy on social media},
  author={Rashidi, Yasmeen and Kapadia, Apu and Nippert-Eng, Christena and Su, Norman Makoto},
  journal={Proceedings of the ACM on human-computer interaction},
  volume={4},
  number={CSCW1},
  pages={1--25},
  year={2020},
  publisher={ACM New York, NY, USA}
}

@article{jhaver2023decentralizing,
    author = {Jhaver, Shagun and Frey, Seth and Zhang, Amy X.}, 
    title = {Decentralizing Platform Power: A Design Space of Multi-level Governance in Online Social Platforms}, 
    year = {2023}, 
    journal = {Social Media + Society}, 
    numpages = {10}, 
    }

@article{juneja2020through,
  title={Through the Looking Glass: Study of Transparency in Reddit's Moderation Practices},
  author={Juneja, Prerna and Rama Subramanian, Deepika and Mitra, Tanushree},
  journal={Proceedings of the ACM on Human-Computer Interaction},
  volume={4},
  number={GROUP},
  pages={1--35},
  year={2020},
  publisher={ACM New York, NY, USA}
}

@article{scheuerman_framework_2021,
	title = {A {Framework} of {Severity} for {Harmful} {Content} {Online}},
	volume = {5},
	url = {https://dl.acm.org/doi/10.1145/3479512},
	doi = {10.1145/3479512},
	number = {CSCW2},
	urldate = {2025-05-12},
	journal = {Proc. ACM Hum.-Comput. Interact.},
	author = {Scheuerman, Morgan Klaus and Jiang, Jialun Aaron and Fiesler, Casey and Brubaker, Jed R.},
	month = oct,
	year = {2021},
	pages = {368:1--368:33},
}

@inproceedings{wohn2020audience,
author = {Wohn, Donghee Yvette and Freeman, Guo}, title = {Audience Management Practices of Live Streamers on Twitch}, year = {2020}, isbn = {9781450379762}, publisher = {Association for Computing Machinery}, address = {New York, NY, USA}, url = {https://doi.org/10.1145/3391614.3393653}, doi = {10.1145/3391614.3393653}, booktitle = {ACM International Conference on Interactive Media Experiences}, pages = {106–116}, numpages = {11}, location = {Cornella, Barcelona, Spain}, series = {IMX '20} }

@article{devito2022transfeminine,
  title={How transfeminine TikTok creators navigate the algorithmic trap of visibility via folk theorization},
  author={DeVito, Michael Ann},
  journal={Proceedings of the ACM on Human-Computer Interaction},
  volume={6},
  number={CSCW2},
  pages={1--31},
  year={2022},
  publisher={ACM New York, NY, USA}
}

@article{cotter2019playing,
  title={Playing the visibility game: How digital influencers and algorithms negotiate influence on Instagram},
  author={Cotter, Kelley},
  journal={New media \& society},
  volume={21},
  number={4},
  pages={895--913},
  year={2019},
  publisher={Sage Publications Sage UK: London, England}
}

@article{verwiebe2024algorithm,
  title={“The algorithm is like a mercurial god”: Exploring content creators’ perception of algorithmic agency on YouTube},
  author={Verwiebe, Roland and Buder, Claudia and Weissmann, Sarah and Osorio-Krauter, Chiara and Philipp, Aaron},
  journal={New Media \& Society},
  pages={14614448241307931},
  year={2024},
  publisher={SAGE Publications Sage UK: London, England}
}

@inproceedings{ma2023multi,
  title={Multi-platform content creation: the configuration of creator ecology through platform prioritization, content synchronization, and audience management},
  author={Ma, Renkai and Gui, Xinning and Kou, Yubo},
  booktitle={Proceedings of the 2023 CHI Conference on Human Factors in Computing Systems},
  pages={1--19},
  year={2023}
}

@article{rahmatulah2024impression,
  title={Impression Management In Building Personal Branding Marco Randy},
  author={Rahmatulah, Dimas and Sunuantari, Manik and Klicek, Tamara},
  journal={INJECT (Interdisciplinary Journal of Communication)},
  volume={9},
  number={2},
  pages={315--328},
  year={2024}
}

@article{hodl2023content,
  title={Content creators between platform control and user autonomy: the role of algorithms and revenue sharing},
  author={H{\"o}dl, Tatjana and Myrach, Thomas},
  journal={Business \& Information Systems Engineering},
  volume={65},
  number={5},
  pages={497--519},
  year={2023},
  publisher={Springer}
}

@article{grimmelmann2015virtues,
  title={The virtues of moderation},
  author={Grimmelmann, James},
  journal={Yale JL \& Tech.},
  volume={17},
  pages={42},
  year={2015},
  publisher={HeinOnline}
}

@article{abidin2016visibility,
  title={Visibility labour: Engaging with Influencers’ fashion brands and\# OOTD advertorial campaigns on Instagram},
  author={Abidin, Crystal},
  journal={Media International Australia},
  volume={161},
  number={1},
  pages={86--100},
  year={2016},
  publisher={SAGE Publications Sage UK: London, England}
}

@article{goffman2002presentation,
  title={The presentation of self in everyday life. 1959},
  author={Goffman, Erving and others},
  journal={Garden City, NY},
  volume={259},
  pages={2002},
  year={2002}
}

@article{cote2017can,
  title={“I can defend myself” women’s strategies for coping with harassment while gaming online},
  author={Cote, Amanda C},
  journal={Games and culture},
  volume={12},
  number={2},
  pages={136--155},
  year={2017},
  publisher={Sage Publications Sage CA: Los Angeles, CA}
}

@inproceedings{heung2024vulnerable,
  title={“Vulnerable, Victimized, and Objectified”: Understanding Ableist Hate and Harassment Experienced by Disabled Content Creators on Social Media},
  author={Heung, Sharon and Jiang, Lucy and Azenkot, Shiri and Vashistha, Aditya},
  booktitle={Proceedings of the 2024 CHI Conference on Human Factors in Computing Systems},
  pages={1--19},
  year={2024}
}

@article{lang2015just,
  title={Just untag it: Exploring the management of undesirable Facebook photos},
  author={Lang, Caroline and Barton, Hannah},
  journal={Computers in Human Behavior},
  volume={43},
  pages={147--155},
  year={2015},
  publisher={Elsevier}
}

@article{litt2012knock,
  title={Knock, knock. Who's there? The imagined audience},
  author={Litt, Eden},
  journal={Journal of broadcasting \& electronic media},
  volume={56},
  number={3},
  pages={330--345},
  year={2012},
  publisher={Taylor \& Francis}
}

@article{chou2022content,
  title={Content creation intention in digital participation based on identity management on Twitch},
  author={Chou, Shih-Wei and Lu, Guan-Ying},
  journal={Behaviour \& Information Technology},
  volume={41},
  number={12},
  pages={2578--2595},
  year={2022},
  publisher={Taylor \& Francis}
}

@article{ellison2006managing,
  title={Managing impressions online: Self-presentation processes in the online dating environment},
  author={Ellison, Nicole and Heino, Rebecca and Gibbs, Jennifer},
  journal={Journal of computer-mediated communication},
  volume={11},
  number={2},
  pages={415--441},
  year={2006},
  publisher={Oxford University Press Oxford, UK}
}

@article{litt2016imagined,
  title={The imagined audience on social network sites},
  author={Litt, Eden and Hargittai, Eszter},
  journal={Social Media+ Society},
  volume={2},
  number={1},
  pages={2056305116633482},
  year={2016},
  publisher={SAGE Publications Sage UK: London, England}
}

@article{duffy2021nested,
  title={The nested precarities of creative labor on social media},
  author={Duffy, Brooke Erin and Pinch, Annika and Sannon, Shruti and Sawey, Megan},
  journal={Social media+ society},
  volume={7},
  number={2},
  pages={20563051211021368},
  year={2021},
  publisher={SAGE Publications Sage UK: London, England}
}

@inproceedings{uttarapong2021harassment,
  title={Harassment experiences of women and LGBTQ live streamers and how they handled negativity},
  author={Uttarapong, Jirassaya and Cai, Jie and Wohn, Donghee Yvette},
  booktitle={Proceedings of the 2021 ACM international conference on interactive media experiences},
  pages={7--19},
  year={2021}
}

@inproceedings{soneji2024feel,
  title={" I feel physically safe but not politically safe": Understanding the Digital Threats and Safety Practices of $\{$OnlyFans$\}$ Creators},
  author={Soneji, Ananta and Hamilton, Vaughn and Doup{\'e}, Adam and McDonald, Allison and Redmiles, Elissa M},
  booktitle={33rd USENIX Security Symposium (USENIX Security 24)},
  pages={1--18},
  year={2024}
}

@article{wu2019agent,
  title={Agent, gatekeeper, drug dealer: How content creators craft algorithmic personas},
  author={Wu, Eva Yiwei and Pedersen, Emily and Salehi, Niloufar},
  journal={Proceedings of the ACM on Human-Computer Interaction},
  volume={3},
  number={CSCW},
  pages={1--27},
  year={2019},
  publisher={ACM New York, NY, USA}
}

@inproceedings{eslami2016first,
  title={First I" like" it, then I hide it: Folk Theories of Social Feeds},
  author={Eslami, Motahhare and Karahalios, Karrie and Sandvig, Christian and Vaccaro, Kristen and Rickman, Aimee and Hamilton, Kevin and Kirlik, Alex},
  booktitle={Proceedings of the 2016 cHI conference on human factors in computing systems},
  pages={2371--2382},
  year={2016}
}

@article{bertaglia2021clout,
  title={Clout chasing for the sake of content monetization: Gaming algorithmic architectures with self-moderation strategies},
  author={Bertaglia, Thales Costa and Dubois, Adrien and Goanta, Catalina},
  journal={Morals \& Machines},
  volume={1},
  number={1},
  pages={22--29},
  year={2021},
  publisher={Nomos}
}

@online{harper2021conspiracy,
  author    = {Jo Harper},
  title     = {How to make money with fake news},
  year      = {2021},
  url       = {https://www.dw.com/en/the-conspiracy-business-how-to-make-money-with-fake-news/a-56660466},
  note      = {Accessed: 2025-09-04},
  publisher = {Deutsche Welle},
}

@article{proudfoot2018saving,
  title={Saving face on Facebook: Privacy concerns, social benefits, and impression management},
  author={Proudfoot, Jeffrey G and Wilson, David and Valacich, Joseph S and Byrd, Michael D},
  journal={Behaviour \& Information Technology},
  volume={37},
  number={1},
  pages={16--37},
  year={2018},
  publisher={Taylor \& Francis}
}

@article{bishop2019managing,
  title={Managing visibility on YouTube through algorithmic gossip},
  author={Bishop, Sophie},
  journal={New media \& society},
  volume={21},
  number={11-12},
  pages={2589--2606},
  year={2019},
  publisher={SAGE Publications Sage UK: London, England}
}

@article{gillett2022safety,
  title={Safety for whom? Investigating how platforms frame and perform safety and harm interventions},
  author={Gillett, Rosalie and Stardust, Zahra and Burgess, Jean},
  journal={Social Media+ Society},
  volume={8},
  number={4},
  pages={20563051221144315},
  year={2022},
  publisher={SAGE Publications Sage UK: London, England}
}

@article{shim2025pin,
  title={The Pin of Shame: Examining Content Creators' Adoption of Pinning Inappropriate Comments as a Moderation Strategy},
  author={Shim, Yunhee and Jhaver, Shagun},
  journal={arXiv preprint arXiv:2505.14844},
  year={2025}
}

@article{ziegele2020not,
  title={Not funny? The effects of factual versus sarcastic journalistic responses to uncivil user comments},
  author={Ziegele, Marc and Jost, Pablo B},
  journal={Communication research},
  volume={47},
  number={6},
  pages={891--920},
  year={2020},
  publisher={Sage Publications Sage CA: Los Angeles, CA}
}

@article{wilson2020hate,
  title={Hate speech on social media: Content moderation in context},
  author={Wilson, Richard Ashby and Land, Molly K},
  journal={Conn. L. Rev.},
  volume={52},
  pages={1029},
  year={2020},
  publisher={HeinOnline}
}

@article{tafesse2023content,
  title={Content creators' participation in the creator economy: Examining the effect of creators’ content sharing frequency on user engagement behavior on digital platforms},
  author={Tafesse, Wondwesen and Dayan, Mumin},
  journal={Journal of Retailing and Consumer Services},
  volume={73},
  pages={103357},
  year={2023},
  publisher={Elsevier}
}

@article{matthews2025supporting,
  title={Supporting the Digital Safety of At-Risk Users: Lessons Learned from 9+ Years of Research and Training},
  author={Matthews, Tara and Bursztein, Elie and Kelley, Patrick Gage and Kissner, Lea and Kramm, Andreas and Oplinger, Andrew and Schou, Andreas and Sleeper, Manya and Somogyi, Stephan and Szostak, Dalila and others},
  journal={ACM Transactions on Computer-Human Interaction},
  volume={32},
  number={3},
  pages={1--39},
  year={2025},
  publisher={ACM New York, NY}
}

@inproceedings{samermit2023millions,
  title={$\{$“Millions$\}$ of people are watching $\{$you”$\}$: Understanding the $\{$Digital-Safety$\}$ Needs and Practices of Creators},
  author={Samermit, Patrawat and Turner, Anna and Kelley, Patrick Gage and Matthews, Tara and Wu, Vanessia and Consolvo, Sunny and Thomas, Kurt},
  booktitle={32nd USENIX Security Symposium (USENIX Security 23)},
  pages={5629--5645},
  year={2023}
}

@article{scheuerman2021framework,
  title={A framework of severity for harmful content online},
  author={Scheuerman, Morgan Klaus and Jiang, Jialun Aaron and Fiesler, Casey and Brubaker, Jed R},
  journal={Proceedings of the ACM on Human-Computer Interaction},
  volume={5},
  number={CSCW2},
  pages={1--33},
  year={2021},
  publisher={ACM New York, NY, USA}
}

@article{kiesler2012regulating,
  title={Regulating behavior in online communities},
  author={Kiesler, Sara and Kraut, Robert and Resnick, Paul and Kittur, Aniket},
  journal={Building successful online communities: Evidence-based social design},
  volume={1},
  pages={4--2},
  year={2012},
  publisher={MIT Press Cambridge, MA}
}

@article{tandoc2020diffusion,
  title={Diffusion of disinformation: How social media users respond to fake news and why},
  author={Tandoc Jr, Edson C and Lim, Darren and Ling, Rich},
  journal={Journalism},
  volume={21},
  number={3},
  pages={381--398},
  year={2020},
  publisher={SAGE Publications Sage UK: London, England}
}

@article{im2022women,
  title={Women's perspectives on harm and justice after online harassment},
  author={Im, Jane and Schoenebeck, Sarita and Iriarte, Marilyn and Grill, Gabriel and Wilkinson, Daricia and Batool, Amna and Alharbi, Rahaf and Funwie, Audrey and Gankhuu, Tergel and Gilbert, Eric and others},
  journal={Proceedings of the ACM on Human-Computer Interaction},
  volume={6},
  number={CSCW2},
  pages={1--23},
  year={2022},
  publisher={ACM New York, NY, USA}
}

@article{cotter2024practical,
  title={Practical knowledge of algorithms: The case of BreadTube},
  author={Cotter, Kelley},
  journal={New Media \& Society},
  volume={26},
  number={4},
  pages={2131--2150},
  year={2024},
  publisher={SAGE Publications Sage UK: London, England}
}

@article{haenlein2020navigating,
  title={Navigating the new era of influencer marketing: How to be successful on Instagram, TikTok, \& Co.},
  author={Haenlein, Michael and Anadol, Ertan and Farnsworth, Tyler and Hugo, Harry and Hunichen, Jess and Welte, Diana},
  journal={California management review},
  volume={63},
  number={1},
  pages={5--25},
  year={2020},
  publisher={SAGE Publications Sage CA: Los Angeles, CA}
}

@inproceedings{choi2023creator,
  title={Creator-friendly algorithms: Behaviors, challenges, and design opportunities in algorithmic platforms},
  author={Choi, Yoonseo and Kang, Eun Jeong and Lee, Min Kyung and Kim, Juho},
  booktitle={Proceedings of the 2023 CHI Conference on Human Factors in Computing Systems},
  pages={1--22},
  year={2023}
}

@article{olsson2022architectures,
  title={Architectures, algorithms \& agency: the information practices of YouTube content creators},
  author={Olsson, Michael},
  year={2022},
  publisher={Philosophische Fakult{\"a}t}
}

@article{glotfelter2019algorithmic,
  title={Algorithmic circulation: how content creators navigate the effects of algorithms on their work},
  author={Glotfelter, Angela},
  journal={Computers and composition},
  volume={54},
  pages={102521},
  year={2019},
  publisher={Elsevier}
}

@article{bolino1999measuring,
  title={Measuring impression management in organizations: A scale development based on the Jones and Pittman taxonomy},
  author={Bolino, Mark C and Turnley, William H},
  journal={Organizational Research Methods},
  volume={2},
  number={2},
  pages={187--206},
  year={1999},
  publisher={Sage Publications Sage CA: Thousand Oaks, CA}
}

@article{weerasinghe2025beyond,
  title={Beyond mute and block: adoption and effectiveness of safety tools in social VR, from ubiquitous harassment to social sculpting},
  author={Weerasinghe, Maheshya and Macdonald, Shaun and Fiani, Cristina and O'Hagan, Joseph and Chollet, Mathieu and McGill, Mark and Khamis, Mohamed},
  journal={IEEE Transactions on Visualization and Computer Graphics},
  year={2025},
  publisher={IEEE}
}

@inproceedings{grober2024chose,
  title={" I chose to fight, be brave, and to deal with it": Threat Experiences and Security Practices of Pakistani Content Creators},
  author={Gr{\"o}ber, Lea and Arshad, Waleed and Goetzen, Angelica and Redmiles, Elissa M and Mustafa, Maryam and Krombholz, Katharina and others},
  booktitle={33rd USENIX Security Symposium (USENIX Security 24)},
  pages={19--36},
  year={2024}
}

@article{stegeman2024strategic,
  title={Strategic invisibility: How creators manage the risks and constraints of online hyper (in) visibility},
  author={Stegeman, Hanne M and Are, Carolina and Poell, Thomas},
  journal={Social Media+ Society},
  volume={10},
  number={2},
  pages={20563051241244674},
  year={2024},
  publisher={SAGE Publications Sage UK: London, England}
}

@article{lee2025managing,
  title={Managing unwanted visibility: how transnational Korean women content creators experience and manage harmful algorithmic visibility on global social media},
  author={Lee, Jeehyun Jenny},
  journal={Information, Communication \& Society},
  pages={1--17},
  year={2025},
  publisher={Taylor \& Francis}
}

@inproceedings{chatzakou2017measuring,
  title={Measuring\# GamerGate: A tale of hate, sexism, and bullying},
  author={Chatzakou, Despoina and Kourtellis, Nicolas and Blackburn, Jeremy and De Cristofaro, Emiliano and Stringhini, Gianluca and Vakali, Athena},
  booktitle={Proceedings of the 26th international conference on world wide web companion},
  pages={1285--1290},
  year={2017}
}

@inproceedings{pater2016characterizations,
  title={Characterizations of online harassment: Comparing policies across social media platforms},
  author={Pater, Jessica A and Kim, Moon K and Mynatt, Elizabeth D and Fiesler, Casey},
  booktitle={Proceedings of the 2016 ACM International Conference on Supporting Group Work},
  pages={369--374},
  year={2016}
}

@inproceedings{thomas2021sok,
  title={Sok: Hate, harassment, and the changing landscape of online abuse},
  author={Thomas, Kurt and Akhawe, Devdatta and Bailey, Michael and Boneh, Dan and Bursztein, Elie and Consolvo, Sunny and Dell, Nicola and Durumeric, Zakir and Kelley, Patrick Gage and Kumar, Deepak and others},
  booktitle={2021 IEEE symposium on security and privacy (SP)},
  pages={247--267},
  year={2021},
  organization={IEEE}
}

@article{eckert2018fighting,
  title={Fighting for recognition: Online abuse of women bloggers in Germany, Switzerland, the United Kingdom, and the United States},
  author={Eckert, Stine},
  journal={New Media \& Society},
  volume={20},
  number={4},
  pages={1282--1302},
  year={2018},
  publisher={SAGE Publications Sage UK: London, England}
}

@article{narayanan2023understanding,
  title={Understanding social media recommendation algorithms},
  author={Narayanan, Arvind},
  year={2023}
}

@inproceedings{wright2017vectors,
  title={Vectors for counterspeech on twitter},
  author={Wright, Lucas and Ruths, Derek and Dillon, Kelly P and Saleem, Haji Mohammad and Benesch, Susan},
  booktitle={Proceedings of the first workshop on abusive language online},
  pages={57--62},
  year={2017}
}

@article{leets2002experiencing,
  title={Experiencing hate speech: Perceptions and responses to anti-semitism and antigay speech},
  author={Leets, Laura},
  journal={Journal of social issues},
  volume={58},
  number={2},
  pages={341--361},
  year={2002},
  publisher={Wiley Online Library}
}

@article{garland2022impact,
  title={Impact and dynamics of hate and counter speech online},
  author={Garland, Joshua and Ghazi-Zahedi, Keyan and Young, Jean-Gabriel and H{\'e}bert-Dufresne, Laurent and Galesic, Mirta},
  journal={EPJ data science},
  volume={11},
  number={1},
  pages={3},
  year={2022},
  publisher={Springer Berlin Heidelberg}
}

@article{obermaier2023ll,
  title={I’ll be there for you? Effects of Islamophobic online hate speech and counter speech on Muslim in-group bystanders’ intention to intervene},
  author={Obermaier, Magdalena and Schmuck, Desir{\'e}e and Saleem, Muniba},
  journal={New Media \& Society},
  volume={25},
  number={9},
  pages={2339--2358},
  year={2023},
  publisher={Sage Publications Sage UK: London, England}
}

@article{sasse2023breaking,
  title={Breaking the silence: Investigating which types of moderation reduce negative effects of sexist social media content},
  author={Sasse, Julia and Grossklags, Jens},
  journal={Proceedings of the ACM on Human-Computer Interaction},
  volume={7},
  number={CSCW2},
  pages={1--26},
  year={2023},
  publisher={ACM New York, NY, USA}
}

@article{ping2024behind,
  title={Behind the Counter: Exploring the Motivations and Barriers of Online Counterspeech Writing},
  author={Ping, Kaike and Kumar, Anisha and Ding, Xiaohan and Rho, Eugenia H},
  journal={ACM Transactions on Computer-Human Interaction},
  year={2024},
  publisher={ACM New York, NY}
}

@incollection{bahador2021countering,
  title={Countering hate speech},
  author={Bahador, Babak},
  booktitle={The Routledge companion to media disinformation and populism},
  pages={507--518},
  year={2021},
  publisher={Routledge}
}

@inproceedings{jhaver2018algorithmic,
  title={Algorithmic anxiety and coping strategies of Airbnb hosts},
  author={Jhaver, Shagun and Karpfen, Yoni and Antin, Judd},
  booktitle={Proceedings of the 2018 CHI conference on human factors in computing systems},
  pages={1--12},
  year={2018}
}

@techreport{madden_smith_2010_reputation,
  author       = {Madden, Mary and Smith, Aaron},
  title        = {Reputation Management and Social Media: How people monitor their identity and search for others online},
  institution  = {Pew Research Center’s Internet \& American Life Project},
  year         = {2010},
  month        = {May},
  day          = {26},
  url          = {https://www.pewresearch.org/internet/2010/05/26/reputation-management-and-social-media/}
}

@article{deandrea2019influence,
  title={The influence of self-generated and third-party claims online: Perceived self-interest as an explanatory mechanism},
  author={DeAndrea, David C and Vendemia, Megan A},
  journal={Journal of Computer-Mediated Communication},
  volume={24},
  number={5},
  pages={223--239},
  year={2019},
  publisher={Oxford University Press}
}

@article{lane2022antecedents,
  title={Antecedents and Effects of Online Third-Party Information on Offline Impressions},
  author={Lane, Brianna L and Cionea, Ioana A and Dunbar, Norah E and Carr, Caleb T},
  journal={Journal of Media Psychology},
  year={2022},
  publisher={Hogrefe Publishing}
}

@inproceedings{klug2021trick,
  title={Trick and please. A mixed-method study on user assumptions about the TikTok algorithm},
  author={Klug, Daniel and Qin, Yiluo and Evans, Morgan and Kaufman, Geoff},
  booktitle={Proceedings of the 13th ACM web science conference 2021},
  pages={84--92},
  year={2021}
}

@article{davidovic2023intervene,
  title={To intervene or not to intervene: Young adults’ views on when and how to intervene in online harassment},
  author={Davidovic, Anna and Talbot, Catherine and Hamilton-Giachritsis, Catherine and Joinson, Adam},
  journal={Journal of Computer-Mediated Communication},
  volume={28},
  number={5},
  pages={zmad027},
  year={2023},
  publisher={Oxford University Press}
}

@article{na2020exploring,
  title={Exploring athlete brand image development on social media: The role of signalling through source credibility},
  author={Na, Sangwon and Kunkel, Thilo and Doyle, Jason},
  journal={European Sport Management Quarterly},
  volume={20},
  number={1},
  pages={88--108},
  year={2020},
  publisher={Taylor \& Francis}
}

@article{tuck2024social,
  title={The social media use scale: Development and validation},
  author={Tuck, Alison B and Thompson, Renee J},
  journal={Assessment},
  volume={31},
  number={3},
  pages={617--636},
  year={2024},
  publisher={Sage Publications Sage CA: Los Angeles, CA}
}

@article{rui2013strategic,
  title={Strategic self-presentation online: A cross-cultural study},
  author={Rui, Jian and Stefanone, Michael A},
  journal={Computers in human behavior},
  volume={29},
  number={1},
  pages={110--118},
  year={2013},
  publisher={Elsevier}
}

@article{calic2023dark,
  title={The dark side of Machiavellian rhetoric: Signaling in reward-based crowdfunding performance},
  author={Calic, Goran and Arseneault, Rene and Ghasemaghaei, Maryam},
  journal={Journal of Business Ethics},
  volume={182},
  number={3},
  pages={875--896},
  year={2023},
  publisher={Springer}
}

@book{NatlAcadScis_2024,
  title        = {Social Media and Adolescent Health},
  editor       = {Wojtowicz, Alexis and Buckley, Gillian J. and Galea, Sandro},
  author       = {{National Academies of Sciences, Engineering, and Medicine; Health and Medicine Division; Board on Population Health and Public Health Practice; Committee on the Impact of Social Media on Adolescent Health}},
  publisher    = {National Academies Press (US)},
  address      = {Washington, DC},
  year         = {2024},
  month        = mar,
  day          = {25},
  note         = {NCBI Bookshelf ID: NBK603437},
  url          = {https://www.ncbi.nlm.nih.gov/books/NBK603437/},
  doi          = {10.17226/27396}
}

@article{yeom2020meta_comments_effects,
  author       = {Yeom, Jeong-yun and Kim, Ryu-won and Jeong, Se-hoon},
  title        = {A Meta-analysis of the Effects of User Comments},
  journal      = {Journal of Communication Research},
  volume       = {57},
  number       = {2},
  pages        = {5--49},
  year         = {2020},
  publisher    = {Institute of Communication Research, Seoul National University},
  url          = {https://s-space.snu.ac.kr/handle/10371/214465},
  note         = {Meta-analysis of domestic comment effect studies; 47 studies analyzed.},
}

@article{rieger2018hate,
  title={Hate and counter-voices in the Internet: Introduction to the special issue},
  author={Rieger, Diana and Schmitt, Josephine B and Frischlich, Lena},
  journal={SCM Studies in Communication and Media},
  volume={7},
  number={4},
  pages={459--472},
  year={2018},
  publisher={Nomos Verlagsgesellschaft mbH \& Co. KG}
}

@inproceedings{schieb2016governing,
  title={Governing hate speech by means of counterspeech on Facebook},
  author={Schieb, Carla and Preuss, Mike},
  booktitle={66th ica annual conference, at fukuoka, japan},
  pages={1--23},
  year={2016}
}

@article{baider2023accountability,
  title={Accountability issues, online covert hate speech, and the efficacy of counter-speech},
  author={Baider, Fabienne},
  journal={Politics and Governance},
  volume={11},
  number={2},
  pages={249--260},
  year={2023},
  publisher={PRT}
}

@article{buerger2022they,
  title={Why they do it: Counterspeech theories of change},
  author={Buerger, Catherine},
  journal={Available at SSRN 4245211},
  year={2022}
}

@inproceedings{scott2023trauma,
  title={Trauma-informed social media: Towards solutions for reducing and healing online harm},
  author={Scott, Carol F and Marcu, Gabriela and Anderson, Riana Elyse and Newman, Mark W and Schoenebeck, Sarita},
  booktitle={Proceedings of the 2023 CHI Conference on Human Factors in Computing Systems},
  pages={1--20},
  year={2023}
}

@article{citron2014addressing,
  title={Addressing cyber harassment: An overview of hate crimes in cyberspace},
  author={Citron, Danielle Keats},
  journal={Case W. Res. JL Tech. \& Internet},
  volume={6},
  pages={1},
  year={2014},
  publisher={HeinOnline}
}

@article{blackwell2017classification,
  title={Classification and its consequences for online harassment: Design insights from heartmob},
  author={Blackwell, Lindsay and Dimond, Jill and Schoenebeck, Sarita and Lampe, Cliff},
  journal={Proceedings of the ACM on human-computer interaction},
  volume={1},
  number={CSCW},
  pages={1--19},
  year={2017},
  publisher={ACM New York, NY, USA}
}

@article{duffy2023platform,
  title={Platform governance at the margins: Social media creators’ experiences with algorithmic (in) visibility},
  author={Duffy, Brooke Erin and Meisner, Colten},
  journal={Media, Culture \& Society},
  volume={45},
  number={2},
  pages={285--304},
  year={2023},
  publisher={SAGE Publications Sage UK: London, England}
}

@article{scolere2018constructing,
  title={Constructing the platform-specific self-brand: The labor of social media promotion},
  author={Scolere, Leah and Pruchniewska, Urszula and Duffy, Brooke Erin},
  journal={Social Media+ Society},
  volume={4},
  number={3},
  pages={2056305118784768},
  year={2018},
  publisher={SAGE Publications Sage UK: London, England}
}

@article{sun2025dynamical,
  title={A dynamical measure of algorithmically infused visibility},
  author={Sun, Shaojing and Liu, Zhiyuan and Waxman, David},
  journal={Royal Society Open Science},
  volume={12},
  number={11},
  year={2025},
  publisher={The Royal Society}
}

@article{guess2023social,
  title={How do social media feed algorithms affect attitudes and behavior in an election campaign?},
  author={Guess, Andrew M and Malhotra, Neil and Pan, Jennifer and Barber{\'a}, Pablo and Allcott, Hunt and Brown, Taylor and Crespo-Tenorio, Adriana and Dimmery, Drew and Freelon, Deen and Gentzkow, Matthew and others},
  journal={Science},
  volume={381},
  number={6656},
  pages={398--404},
  year={2023},
  publisher={American Association for the Advancement of Science}
}

@article{campbell2020more,
  title={More than meets the eye: The functional components underlying influencer marketing},
  author={Campbell, Colin and Farrell, Justine Rapp},
  journal={Business horizons},
  volume={63},
  number={4},
  pages={469--479},
  year={2020},
  publisher={Elsevier}
}

@article{davis1989perceived,
  title={Perceived usefulness, perceived ease of use, and user acceptance of information technology},
  author={Davis, Fred D},
  journal={MIS quarterly},
  volume={13},
  number={3},
  pages={319--340},
  year={1989},
  publisher={Management Information Systems Research Center, University of Minnesota}
}

@inproceedings{kang2025understanding,
  title={Understanding Content Creators' Struggles and Expectations of AI in Direct Messaging},
  author={Kang, Eun Jeong and Chen, Jingruo and Fussell, Susan R},
  booktitle={Proceedings of the Extended Abstracts of the CHI Conference on Human Factors in Computing Systems},
  pages={1--8},
  year={2025}
}

@article{echauri2026algorithmic,
  title={Algorithmic-driven exposure: hyper-saturation and ever-expansion in the content creator ecosystem},
  author={Echauri, Guillermo},
  journal={Celebrity Studies},
  pages={1--15},
  year={2026},
  publisher={Taylor \& Francis}
}

@article{huang2025empowering,
  title={Empowering Creators in the Fight Against Online Hate: A Qualitative Exploration of AI-Mediated Counterspeech Tools},
  author={Huang, Phoebe Yiqing and Deng, Jiaming and Yang, Yingchen and Williams, Spencer},
  journal={Proceedings of the ACM on Human-Computer Interaction},
  volume={9},
  number={7},
  pages={1--27},
  year={2025},
  publisher={ACM New York, NY, USA}
}

@book{roberts2014behind,
  title={Behind the screen: The hidden digital labor of commercial content moderation},
  author={Roberts, Sarah T},
  year={2014},
  publisher={University of Illinois at Urbana-Champaign}
}

@article{marwick2015instafame,
  title={Instafame: Luxury selfies in the attention economy},
  author={Marwick, Alice E},
  journal={Public culture},
  volume={27},
  number={1},
  pages={137--160},
  year={2015},
  publisher={Duke University Press}
}

@article{goldhaber1997attention,
  title={The attention economy and the net},
  author={Goldhaber, Michael H},
  journal={First monday},
  year={1997}
}

@article{kopf2020rewarding,
  title={“Rewarding good creators”: Corporate social media discourse on monetization schemes for content creators},
  author={Kopf, Susanne},
  journal={Social Media+ Society},
  volume={6},
  number={4},
  pages={2056305120969877},
  year={2020},
  publisher={Sage Publications Sage UK: London, England}
}

@article{kihlstrom2021ecological,
  title={Ecological validity and “ecological validity”},
  author={Kihlstrom, John F},
  journal={Perspectives on Psychological Science},
  volume={16},
  number={2},
  pages={466--471},
  year={2021},
  publisher={Sage Publications Sage CA: Los Angeles, CA}
}

@article{kaiser2009group,
  title={Group identification moderates attitudes toward ingroup members who confront discrimination},
  author={Kaiser, Cheryl R and Hagiwara, Nao and Malahy, Lori W and Wilkins, Clara L},
  journal={Journal of Experimental Social Psychology},
  volume={45},
  number={4},
  pages={770--777},
  year={2009},
  publisher={Elsevier}
}

@book{benoit2014accounts,
  title={Accounts, excuses, and apologies: Image repair theory and research},
  author={Benoit, William L},
  year={2014},
  publisher={Suny Press}
}

@article{sukmayadi2024constructing,
  title={Constructing fame: A phenomenological study of online impression management among Indonesian TikTok celebrities},
  author={Sukmayadi, Vidi and Darmawangsa, Dante and Ayub, Suffian Hadi and Fadhila, Sarah Annisa},
  journal={The Qualitative Report},
  volume={29},
  number={6},
  pages={1727--1741},
  year={2024},
  publisher={The Qualitative Report}
}

@inproceedings{rashed2026if,
  title={What If Moderation Didn’t Mean Suppression? A Case for Personalized Content Transformation},
  author={Rashed, Rayhan and Jahanbakhsh, Farnaz},
  booktitle={Proceedings of the 2026 CHI Conference on Human Factors in Computing Systems},
  pages={1--22},
  year={2026}
}

@article{murphy2026can,
  title={“It can feel uncomfortable to say something, but it's for the best of the group”: Examining Moderator Practices in Online Pregnancy Loss Communities},
  author={Murphy, Stephanie and Peelo Dennehy, Doireann and Morrissey, Kellie and McCarthy, John and Foley, Sarah},
  journal={Proceedings of the ACM on Human-Computer Interaction},
  volume={10},
  number={2},
  pages={1--34},
  year={2026},
  publisher={ACM New York, NY}
}

@inproceedings{spiel2019how,
  author    = {Spiel, Katta and Haimson, Oliver L. and Lottridge, Danielle},
  title     = {How to Do Better with Gender on Surveys: A Guide for HCI Researchers},
  booktitle = {Extended Abstracts of the 2019 CHI Conference on Human Factors in Computing Systems},
  pages     = {1--8},
  year      = {2019},
  doi       = {10.1145/3290607.3310439}
}

@inproceedings{blackwell2019harassment,
  title={Harassment in social VR: Implications for design},
  author={Blackwell, Lindsay and Ellison, Nicole and Elliott-Deflo, Natasha and Schwartz, Raz},
  booktitle={2019 IEEE conference on virtual reality and 3D user interfaces (VR)},
  pages={854--855},
  year={2019},
  organization={IEEE}
}

@inproceedings{andalibi2017sensitive,
  title={Sensitive self-disclosures, responses, and social support on Instagram: The case of\# depression},
  author={Andalibi, Nazanin and Ozturk, Pinar and Forte, Andrea},
  booktitle={Proceedings of the 2017 ACM conference on computer supported cooperative work and social computing},
  pages={1485--1500},
  year={2017}
}

@article{mariotto2026understanding,
  title={Understanding Online Hate Toward Sexual and Gender Minorities: A Systematic Review},
  author={Mariotto, Michela and Costa, Sara and Di Brango, Noemi and Corbelli, Giuseppe and Verbena, Serena and Palladino, Benedetta Emanuela and Zuffian{\`o}, Antonio and Ioverno, Salvatore and others},
  journal={TRAUMA, VIOLENCE \& ABUSE},
  pages={0--19},
  year={2026}
}

@article{keum2024racial,
  title={Racial hate at the intersection of online and offline worlds: The joint impact of online and offline racism on the mental health of racially minoritized individuals},
  author={Keum, Brian TaeHyuk and Wong, Michele J and Sanders, India},
  journal={Journal of interpersonal violence},
  volume={39},
  number={11-12},
  pages={2487--2506},
  year={2024},
  publisher={Sage Publications Sage CA: Los Angeles, CA}
}

\clearpage

\appendix
\section{Survey Sample Description}


\begin{table}[ht]
\centering
\caption{Demographics of Survey Respondents.}
\label{sec:appendix_survey_sample1}
\resizebox{0.70\textwidth}{!}{%
\label{table:Demo_stat}
\small
\begin{tabular}{lll}
\toprule
 \textbf{Demographic Factor} & \textbf{Category} & \textbf{Number (\%)} \\
\midrule
Gender
 & \quad Woman & 327 (56\%) \\
 & \quad Man & 248 (42.5\%) \\
 & \quad Non-binary & 6 (1\%) \\
 & \quad Prefer not to answer & 3 (0.5\%) \\
\midrule

Age
 & \quad 18-24  & 136 (23\%) \\
 & \quad 25-29  & 159 (27.2\%) \\
 & \quad 30-34  & 131 (22.4\%) \\
 & \quad 35-39  & 57 (9.8\%) \\
 & \quad 40-44  & 43 (7.4\%) \\
 & \quad 45-49  & 25 (4.3\%) \\
 & \quad 50-54  & 15 (2.6 \%) \\
 & \quad 55-59  & 12 (2.1\%) \\
 & \quad 60-54  & 4 (0.7\%) \\
 & \quad 65 above  & 2 (0.3\%) \\
\midrule

Ethnicity
 & \quad Black or African American & 228 (39\%) \\
  & \quad White & 222 (38\%) \\
 & \quad Asian & 83 (14.2\%) \\
 & \quad Latino/Hispanic & 15 (2.6\%) \\
  & \quad Mixed race & 14 (2.4\%) \\
   & \quad Prefer not to answer & 10 (1.7\%) \\
  & \quad Middle Eastern or North African & 9 (1.5\%) \\
   & \quad American Indian or Alaska Native & 3 (0.5\%) \\

\midrule
 Educational Attainment
 & \quad High school or less & 60 (10.3\%) \\
 & \quad Some college or associate degree & 90 (15.5\%) \\
 & \quad Bachelor's degree & 298 (51\%) \\
 & \quad Master's degree or equivalent & 111 (19\%) \\
 & \quad Doctorate degree & 23 (3.9\%) \\
 & \quad Prefer not to answer & 2 (0.3\%) \\

\bottomrule
\end{tabular}
}
\label{table:Demo_stat}
\end{table}

\begin{table}[ht]
\centering
\caption{Creator Characteristics of Survey Respondents}
\label{sec:appendix_survey_sample2}
\resizebox{0.7\textwidth}{!}{%
\label{table:Demo_stat}
\small
\begin{tabular}{lll}
\toprule
 \textbf{Category} & \textbf{item} & \textbf{Number (\%)} \\
\midrule

Content creation platform

 & \quad TikTok & 422 ( 72.3\%) \\
 & \quad Instagram & 406 (69.5\%) \\
 & \quad YouTube & 340 (58.2\%) \\
 & \quad Facebook & 317 ( 54.3 \%) \\
 & \quad Twitch & 47   ( 8\%) \\
 & \quad Others & 48   ( 8.2\%) \\
 & \quad Prefer not to answer & 2 (0.3\%) \\

 \midrule
Content creation topic
 & \quad About me and private life & 209 (35.8\%) \\
 & \quad Beauty and cosmetics & 138 (23.6\%) \\
 & \quad Business and management & 68 (11.6\%) \\
 & \quad Comedy and entertainment & 157 (26.9\%) \\
 & \quad Do it yourself & 71 (12.2\%) \\
 & \quad Esoteric topics & 15 (2.6\%) \\
 & \quad Fashion & 118 (20.2\%) \\
 & \quad Food and cooking & 127 (21.7\%) \\
 & \quad Gaming and games & 120 (20.5\%) \\
 & \quad Health and fitness & 1 (0.2\%) \\
 & \quad Lifestyle & 187 (32\%) \\
 & \quad Music and art & 99 (17\%) \\
 & \quad News and politics & 50 (8.6\%) \\ 
 & \quad Pets and animals & 35 (6\%) \\
 & \quad Psychology and mindfulness & 48 (8.2\%) \\
 & \quad Science and technology & 75 (12.8\%) \\
 & \quad Travel and outdoors & 112 (19.2\%) \\
 & \quad Others & 42 (0.8\%) \\

\midrule
Number of followers
& \quad Less than 1,000 & 103 (17.6\%) \\
& \quad 1,001-5,000 & 189 (32.4\%) \\
& \quad 5,001-10,000 & 102 (17.5\%) \\
& \quad 10,001-50,000 & 126 (21.6\%) \\
& \quad 50,001-100,000 & 35 (6\%) \\
& \quad More than 100,000 & 24 (4.1\%) \\
& \quad Prefer not to answer & 5 (0.9\%) \\

\bottomrule
\end{tabular}
}

\vspace{2mm}
\begin{minipage}{0.7\linewidth}
\footnotesize
\textit{Note.} Multiple responses were allowed. ``Others'' includes uncategorized responses.
\end{minipage}
\end{table}

\vspace{5em}

\end{document}